\documentclass[11pt]{article}   	

\usepackage{graphicx}				
\usepackage{amssymb}

\usepackage{footmisc,authblk}
\usepackage{subfig}
\usepackage{natbib,url}
\usepackage{lineno,amsmath}

\newcommand*{\be}{\begin{equation}}
\newcommand*{\ee}{\end{equation}}

\def\begineq{\begin{equation}}
\def\endeq{\end{equation}}

\def\begineqn{\begin{equation*}}
\def\endeqn{\end{equation*}}

\def\beginar{\begin{eqnarray}}
\def\endar{\end{eqnarray}}
\def\beginarn{\begin{eqnarray*}}
\def\endarn{\end{eqnarray*}}

\def\lb{\left ( }
\def\rb{\right ) }
\def\lsq{\left [ }
\def\rsq{\right ] }

\def\ep{\epsilon}

\def\Rat{\widetilde{Ra}}
\def\Rmt{\widetilde{Rm}}
\def\Pmt{\widetilde{Pm}}

\def\ub{\mathbf{u}}
\def\He{\overline{\mathcal{H}}}
\def\Bb{\mathbf{B}}
\def\ubp{\mathbf{u}^{\prime}}
\def\bbp{\mathbf{b}^{\prime}}

\def\mBb{\overline{\bf B}}

\def\mth{\overline{\vartheta}}
\def\pth{\vartheta^{\prime}}

\def\mBx{\overline{B}_x}
\def\mBy{\overline{B}_y}

\def\pbx{b^{\prime}_x}
\def\pby{b^{\prime}_y}
\def\pbz{b^{\prime}_z}

\def\bz{b_z}
\def\pjb{\mathbf{j}^{\prime}}
\def\pjx{j^{\prime}_x}
\def\pjy{j^{\prime}_y}
\def\pjz{j^{\prime}_z}

\def\dt{{\partial_{T}}}
\def\dtau{{\partial_{\tau}}}
\def\dta{{\partial_\tau}}

\def\dsx{{\partial_x}}
\def\dsy{{\partial_y}}

\def\dsyy{{\partial_{yy}}}
\def\dsxx{{\partial_{xx}}}
\def\dsxy{{\partial_{xy}}}
\def\dst{{\partial_t}}

\def\dsz{{\partial_z}}

\def\dz{{\partial_Z}}
\def\dzt{{\partial^2_Z}}

\def\hz{{\bf\widehat z}}

\def\hy{{\bf\widehat y}}
\def\hx{{\bf\widehat x}}

\def\litx{{\bf x}}

\def\emf{\overline{\boldsymbol{\mathcal{E}}}}
\def\emfx{\mathcal{E}_x}
\def\emfy{\mathcal{E}_y}

\def\md{D^\perp_t}
\def\div{{\nabla \cdot}}

\def\lp{{\nabla_\perp^2}}
\def\lpi{{\nabla_\perp^{-2}}}

\title{Convection-driven kinematic dynamos at low Rossby and magnetic Prandtl numbers: single mode solutions}
\author[1]{Michael A. Calkins} 
\author[2]{Keith Julien}
\author[3]{Steven M. Tobias}
\author[4]{Jonathan M. Aurnou}
\author[2]{Philippe Marti}

\affil[1]{Department of Physics, University of Colorado, Boulder, CO  80309, USA}
\affil[2]{Department of Applied Mathematics, University of Colorado, Boulder, CO  80309, USA}
\affil[3]{Department of Applied Mathematics, University of Leeds, Leeds, UK LS2 9JT}
\affil[4]{Department of Earth, Planetary and Space Sciences, University of California, Los Angeles, CA  90095, USA}

\date{}							

\begin{document}
\maketitle
\begin{abstract}
The onset of dynamo action is investigated within the context of a newly developed low Rossby, low magnetic Prandtl number, convection-driven dynamo model.  This multiscale model represents an asymptotically exact form of an $\alpha^2$ mean field dynamo model in which the small-scale convection is represented explicitly by finite amplitude, single mode solutions.  Both steady and oscillatory convection are considered for a variety of horizontal planforms.  The kinetic helicity is observed to be a monotonically increasing function of the Rayleigh number. As a result, very small magnetic Prandtl number dynamos can be found for sufficiently large Rayleigh numbers.  All dynamos are found to be oscillatory with an oscillation frequency that increases as the strength of the convection is increased and the magnetic Prandtl number is reduced.  Kinematic dynamo action is strongly controlled by the profile of the helicity; single mode solutions which exhibit boundary layer behavior in the helicity show a decrease in the efficiency of dynamo action due to the enhancement of magnetic diffusion in the boundary layer regions. For a given value of the Rayleigh number, lower magnetic Prandtl number dynamos are excited for the case of oscillatory convection in comparison to steady convection.  With regards to planetary dynamos, these results suggest that the low magnetic Prandtl number dynamos typical of liquid metals are more easily driven by thermal convection than by compositional convection.
\end{abstract}

\section{Introduction}


Turbulent rotating convection is thought to be the primary mechanism for sustaining the observed large-scale magnetic fields of stars and planets via the dynamo mechanism \citep{eP55,hM78b,cJ11b}.  Indeed, the majority of planetary magnetic fields are predominantly aligned with the rotation axis of the planet and suggests that planetary rotation, through the action of the Coriolis force, is an important ingredient in generating these fields \citep{gS11}.   Two non-dimensional parameters that control the dynamics of rotating convection are the Rossby and Ekman numbers defined by, respectively,
\be
Ro = \frac{U}{2\Omega L}, \quad Ek = \frac{\nu}{2\Omega L^2} ,
\ee
where $U$ is a characteristic velocity scale, $L$ is a characteristic length scale, $\nu$ is the kinematic viscosity of the fluid, and $\Omega$ is the planetary rotation rate.  Geostrophically balanced convection, in which the Coriolis and pressure gradient forces are of equal magnitude approximately, is characterized by the limit $(Ro, Ek) \rightarrow 0$.  

Despite its relevance to geophysical and astrophysical systems, the turbulent geostrophic convection regime is intrinsically difficult to study via both direct numerical simulations (DNS) and laboratory studies \citep{sS10,uC11,cJ11b,rS13,bF14,sS14,rE14,jmA15,jC15,cG15}.  In the former case, the governing equations are numerically stiff, requiring vast computational resources to resolve the disparate spatiotemporal scales that characterize these systems.  For the laboratory, accessing the $(Ro, Ek) \rightarrow 0$ cannot be done owing to mechanical constraints of experimental hardware.  Such difficulties are common across many areas of physics and have motivated the development of asymptotic models that rigorously simplify the governing equations.   With regards to dynamos, the \cite{sC72} and \cite{aS74} first showed that an asympotically reduced, weakly nonlinear dynamo model can be developed in the geostrophic limit for the classical Rayleigh-B\'enard, or plane layer geometry.  Their work focused on fluids characterized by an order one magnetic Prandtl number,
\be
 Pm = \nu/\eta, 
 \ee
where $\eta$ is the magnetic diffusivity, and showed conclusively that rapidly rotating convective flows act as an efficient generator of magnetic field.

Understanding the differences between dynamos with both large and small $Pm$ continues to be a fundamental aspect of dynamo theory \citep{aS04,yP05,nS06,aI07}.  Whereas galactic plasmas are thought to be characterized by $Pm \gg 1$, the $Pm \ll 1$ limit is representative of stellar convection zones, planetary cores and laboratory experiments \citep{rM07,sC14,eK15,aR15}.  DNS investigations are limited to relatively small values of the Reynolds number, $Re = UL/\nu$.  Depending upon the particular form of fluid motion, order one and greater magnetic Reynolds numbers, $Rm = Pm Re$, are required for the onset of dynamo action \citep[e.g.][]{hM78b}.  Fluids with $Pm \ge 1 $ then have the advantage that dynamos can be simulated with relatively small Reynolds numbers. However, the relevance of such studies to the large Reynolds number flows characteristic of nearly all natural dynamos remains an open question \citep{aB12,pR13,jmA15}.  Small $Pm$ dynamos are therefore more computationally demanding, given that larger $Re$ and thus larger numerical resolution is required to sustain magnetic field growth. For such dynamos ohmic dissipation occurs on length scales that are within the inertial range of the turbulence and hinders dynamo action at moderate $Rm$ \citep{sT11}.  Increases in computational power are nevertheless allowing DNS investigations to reduce $Pm$, with such studies finding quite different behavior in comparison to $Pm \ge 1$ dynamos \citep[e.g.][]{cG15}.   However, convection-driven dynamos with geo- and astrophysically realistic values of $Pm$ have so far been unattainable with DNS.   


Recent work has shown that the geostrophic dynamo model of \cite{sC72} can be rigorously generalized to the more geo- and astrophysically relevant case of small magnetic Prandtl number and strongly nonlinear motions \citep{mC15b}; we refer to this model as the quasi-geostrophic dynamo model (QGDM).  In the present work we investigate the kinematic dynamo problem for the QGDM.  We focus on a simplified class of finite amplitude rotating convective solutions, first investigated by \cite{aB94} and \citep{kJ97}, that satisfy the nonlinear reduced system of governing hydrodynamic equations exactly \citep{kJ99b}.  We refer to these solutions as `single mode' since they are characterized by a single horizontal spatial wavenumber and an analytical horizontal spatial structure, or planform.  Although the solutions are specialized given that advective nonlinearities vanish identically, they allow us to approach the dynamo problem with relative ease and will provide a useful point of comparison for future numerical simulations of the QGDM. Moreover, turbulent solutions are often thought of as a superposition of these modes.  As shown below, the simplified mathematical structure of the single mode solutions results in magnetic induction equations that show explicitly the mechanism for dynamo action in rapidly rotating convection.

In section \ref{S:QGDM} the QGDM is summarized and single mode theory is reviewed in section \ref{S:smode}.  Numerical results and concluding remarks are presented in sections \ref{S:results} and \ref{S:conclude}, respectively.

\section{Governing Equations}
\label{S:govern}

\subsection{Summary of the quasi-geostrophic dynamo model (QGDM)}
\label{S:QGDM}

Here we briefly summarize the QGDM for the particular case of $Pm \ll 1$ and no large-scale horizontal modulation.  The latter simplification implies that both the large-scale (mean) velocity and the vertical component of the mean magnetic field are identically zero.  For a more detailed discussion of the derivation see \citep{mC15b}.  We consider a rotating layer of Boussinesq fluid of dimensional depth $H$, heated from below and cooled from above with temperature difference $\Delta T$.  The constant rotation and gravity vectors are given by $\mathbf{\Omega} = \Omega \hz$ and $\mathbf{g} = - g\hz$, respectively, and the vertical unit vector $\hz$ points upwards.  The thermal and electrical properties of the fluid are quantified by thermal diffusivity $\kappa$ and magnetic diffusivity $\eta$.  In what follows we non-dimensionalize the equations using the small-scale, dimensional viscous diffusion time $t^* = L^2/\nu$.

The QGDM is derived by assuming that the convection is geostrophically balanced in the sense that $\ep = Ek^{1/3} \rightarrow 0$.  In this limit it is well-known that convection is highly anisotropic with aspect ratio $A = H/L = \ep^{-1} \gg 1$, where the horizontal scale of convection is $L = H Ek^{1/3}$ \citep{sC61}.  Multiple scales are then employed in space and time such that the differentials become 
\begin{gather}
\dst  \rightarrow  \dst + \ep^{3/2} \dta + \ep^2 \dt , \\
\dsz  \rightarrow  \dsz + \ep \dz  , \\
\dsx \rightarrow \dsx, \quad \dsy \rightarrow \dsy ,
\end{gather}
where $Z=\ep z$ is the large-scale vertical coordinate, $t$ is the `fast' convective timescale, $\tau = \ep^{3/2} t$ is the slow evolution timescale of the mean magnetic field, and $T = \ep^2 t$ is the slow evolution timescale of the mean temperature.   In \cite{mC15b} it was shown that to allow for a time-dependent dynamo, the mean magnetic timescale was required to be $\tau = \ep^{3/2} t$ for the particular case of $(\ep,Pm) \ll 1$.  The small-scale (fast) independent variables are $(\litx,t)$ and the large-scale (slow) independent variables $(Z,\tau,T)$.  We denote $\ub$ as the velocity vector, $\vartheta$ is the temperature, $p$ is the pressure, and $\Bb$ is the magnetic field vector. Convective motions occur over the large-scale vertical coordinate $Z$, and the Proudman-Taylor theorem is satisfied over the small-scale vertical coordinate.  


Each of the dependent variables are decomposed into mean and fluctuating terms according to $f = \overline{f} + f'$, where the mean is defined by
\begineq
\overline{f}(Z,\tau,T)  = \lim_{t', \mathcal{V} \rightarrow \infty} \, \frac{1}{t' \mathcal{V}} \int_{t',\mathcal{V}} f(\litx,Z,t,\tau,T) \, d \mathbf{x} \, dt , \quad \overline{f'} \equiv 0,
\endeq
and $\mathcal{V}$ is the small-scale fluid volume.  Each variable is expanded in a power series according to
\be
\begin{split}
f(\litx,Z,t,\tau,T)  = \overline{f}_0(Z,\tau,T) + f'_0(\litx,Z,t,\tau,T) + \\  \ep^{1/2} \lsq \overline{f}_{1/2}(Z,\tau,T) + f'_{1/2}(\litx,Z,t,\tau,T) \rsq + \\ \ep \lsq \overline{f}_{1}(Z,\tau,T) + f'_{1}(\litx,Z,t,\tau,T) \rsq + O(\ep^{3/2}) . \label{E:expand}
\end{split}
\ee

The QGDM is derived by employing the above expansion for each variable and separating the governing equations into mean and fluctuating components.  This procedure leads to geostrophically balanced, horizontally divergence-free fluctuating momentum dynamics according to
\be
\hz \times \ubp_0 = - \nabla_\perp p'_1 , \quad \nabla_\perp \cdot \ubp_0 = 0,
\ee
where $\nabla_\perp = (\dsx,\dsy,0)$.  We can then define the geostrophic streamfunction $\psi_0' \equiv p'_1$ such that $\ubp_{0,\perp} = (u_0', v_0') = - \nabla \times \psi_0' \hz$.  The vertical vorticity is then given by $\zeta_0' = \hz \cdot \nabla \times \ubp_0 = \lp \psi_0'$.  Ageostrophic motions $\ub_{1}'$ provide the source for vortex stretching via mass conservation at $O(\ep)$,
\be
\div \ubp_1 + \dz w_0' = 0.
\ee




Hereafter we omit the asymptotic ordering subscripts for notational simplicity. The reduced governing system of equations then consist of the vertical vorticity, vertical momentum, fluctuating heat, mean heat, mean magnetic field and fluctuating magnetic field equations, respectively \citep[cf.][]{mC15b}, 
\begin{gather}
\md \lp \psi' - \dz w' = \mBb \cdot \nabla_\perp \pjz + \nabla_\perp^4 \psi', \label{E:fvort} \\
\md w' +  \dz \psi' = \frac{\Rat}{Pr}  \pth  + \mBb \cdot \nabla_\perp \pbz + \lp w', \label{E:fmom} \\
 \md \pth +  w' \dz \mth = \frac{1}{Pr} \lp \pth , \label{E:fheat} \\
 \dt \mth + \dz \overline{\lb w' \pth \rb} =  \frac{1}{Pr} \dzt \mth , \label{E:mheat} \\
  \partial_\tau \mBb =  \hz \times \dz \emf  + \frac{1}{\Pmt} \dzt \mBb, \label{E:minduc1}\\
  0 = \mBb \cdot \nabla_\perp \ubp + \frac{1}{\Pmt} \lp  \bbp \label{E:finduc1},
\end{gather}
where $\md = \dst + \ubp \cdot \nabla_\perp$.  Here the mean and fluctuating components of the temperature and magnetic field vector are related by $\vartheta = \mth + \ep \pth$ and $\Bb = \mBb + \ep^{1/2} \bbp$, respectively.  The components of the mean and fluctuating magnetic field vectors are denoted by $\mBb = (\mBx,\mBy,0)$ and $\bbp = (\pbx,\pby,\pbz)$, and the corresponding fluctuating current density is $\pjb = (\pjx,\pjy,\pjz) = (\dsy \pbz,-\dsx \pbz, \dsx \pby - \dsy \pbx)$.  The mean electromotive force (emf) present in equation \eqref{E:minduc1} is denoted by $\emf = \overline{\lb \ubp \times \bbp \rb}$.

The non-dimensional parameters appearing above are the reduced Rayleigh number, the reduced magnetic Prandtl number and the Prandtl number defined by
\be
\Rat = Ra Ek^{4/3}, \quad \Pmt = \ep^{-1/2} Pm  , \quad Pr = \frac{\nu}{\kappa}, 
\ee
where the traditional Rayleigh number is 
\be
Ra = \frac{g \alpha \Delta T H^3}{\nu \kappa} .
\ee


In the present model we have $Pr=O(1)$ in the asymptotic sense.  For geostrophy to hold we require that $Pr \gg Ek$, which is thought to hold in most fluids of geo- and astrophysical interest.  When $Pr = O(Ek)$ the convective motions vary on the same timescale as the (dimensional) background rotation timescale $\Omega^{-1}$, and are therefore not geostrophically balanced \citep[cf.][]{kZ97b}.

We note that the $Pm = \ep^{1/2}\Pmt$ relation implies that the present model is implicitly low magnetic Prandtl number.  The small-scale reduced magnetic Reynolds number is given by $\Rmt = \Pmt Re = O(1)$, where the small-scale Reynolds number $Re=O(1)$ is a non-dimensional measure of the convective speed.  Importantly, both the large-scale magnetic Reynolds number and the large-scale Reynolds number based on the depth of the fluid layer are large, 
\be
Rm_H = \ep^{-1/2} \Rmt \gg 1, \quad Re_H = \ep^{-1} Re \gg 1.
\ee
Moreover, it was shown in \cite{mC15b} that the ratio of the magnetic energy to the kinetic energy is also large for the $Pm \ll 1$ model,
\be
M = \frac{B^2}{\rho \mu U^2} = \ep^{-1/2} \gg 1,
\ee
where $B$ is the magnitude of the large-scale magnetic field.

\cite{mC15b} also derived a $Pm=O(1)$ QGDM that possesses many distinct features in comparison to the low $Pm$ model discussed in the present work. There it was shown that when $Pm=O(1)$ both the mean and fluctuating magnetic field components are order one in magnitude, $M=O(1)$, the mean magnetic field evolves on a faster timescale in comparison to the $Pm\ll 1$ case, and the Lorentz force involves fluctuating-fluctuating Reynolds stresses in addition to the mean-fluctuating stresses present in equations \eqref{E:fvort}-\eqref{E:fmom}.  Given that $Pm=O(1)$ is the parameter regime accessible to most DNS investigations, it will be of use to also study the properties of the $Pm=O(1)$ QGDM in future work.

Impenetrable, constant temperature boundary conditions are used for all the reported results,
\be
w' = 0 \quad \textnormal{at} \quad Z = 0, 1,
\ee
\be
\mth = 1\quad \textnormal{at} \quad Z = 0, \quad \textnormal{and} \quad 
\mth = 0\quad \textnormal{at} \quad Z = 1 .
\ee
We note that the thermal boundary conditions become unimportant in the absence of large-scale horizontal modulations since the convective solutions for constant temperature and constant heat flux boundary conditions are asymptotically equivalent in the limit $\ep \rightarrow 0$ \citep{mC15c}.

The magnetic boundary conditions are perfectly conducting for the mean field,
\be
\dz \mBb = 0 \quad \textnormal{at} \quad Z = 0, 1 .
\ee
Examination of equation \eqref{E:finduc1} shows that the fluctuating magnetic field obeys the same boundary conditions as the mean field.  Recent work has shown that the kinematic dynamo problem is independent of the magnetic boundary conditions in the sense that both perfectly conducting and perfectly insulating magnetic boundary conditions yield identical results for reversible flows, i.e.~$\ubp\rightarrow-\ubp$ \citep{bF13c}.  In the present work this property is satisfied if we also take the thermal perturbations to be reversible in the sense $\pth \rightarrow -\pth$.

\subsection{Single mode theory}
\label{S:smode}

In the present section we review the hydrodynamic single mode theory developed by \cite{aB94} and utilized by \cite{kJ99b} in their investigation of single mode solutions of the quasi-geostrophic convection equations developed by \cite{kJ98a}.  See also the single mode studies of \citep{pB94} and \citep{kJ99c,pM99} for the non-rotating magnetoconvection cases, respectively.  For the kinematic dynamo problem, the Lorentz force is neglected and the vorticity, vertical momentum, fluctuating heat and mean heat equations (\eqref{E:fvort}-\eqref{E:fheat}) then decouple from induction processes to become
\begin{gather}
\md \lp \psi' - \dz w' = \nabla_\perp^4 \psi', \label{E:fvort2} \\
\md w' +  \dz \psi' = \frac{\Rat}{Pr}  \pth  + \lp w', \\
 \md \pth +  w' \dz \mth = \frac{1}{Pr} \lp \pth \label{E:fheat2}, \\
 \dt \mth + \dz \overline{\lb w' \pth \rb} =  \frac{1}{Pr} \dzt \mth . \label{E:mheat2} 
\end{gather}

Here we are concerned with solutions that lead to well-defined steady or time-periodic single-frequency states.  As such, the averaging operator present in the mean heat equation yields a mean temperature for which $\dt \mth\equiv 0$.  For these cases we can integrate equation \eqref{E:mheat} directly to yield 
\be
Pr \overline{\lb w' \theta \rb} = \dz \mth + Nu,
\ee
valid at all vertical levels.  $Nu$ is the non-dimensional measure of the global heat transport, or Nusselt number, defined by 
\be
Nu = - \dz \mth |_{Z=0} = - \dz \mth |_{Z=1} = 1 + Pr \langle \overline{w' \pth} \rangle ,
\ee
where the angled brackets denote an average over the large-scale vertical coordinate $Z$.

The solutions for the fluctuating streamfunction, vertical velocity and temperature take the form 
\be 
\lb \psi', w', \pth \rb = \frac{1}{2}\lsq \Psi(Z),W(Z),\Theta(Z) \rsq h(x,y) e^{i \omega t} + c.c. , \label{E:ansatz}
\ee
 in which $h$ satisfies the planform equation
\be
\lp h = - k^2 h, \quad \overline{h^2} \equiv 1, \label{E:plan}
\ee
and $\omega$ is the nonlinear oscillation frequency.  For steady convection ($Pr \gtrsim 0.68$) we simply set $\omega = 0$ \citep{sC61}.  Upon substitution of the ansatz given by equation \eqref{E:ansatz}, all nonlinear advection terms appearing in equations \eqref{E:fvort2}-\eqref{E:fheat2} vanish identically so that the only remaining nonlinearity is the vertical convective flux appearing on the left-hand side of the mean heat equation \eqref{E:mheat2}.  The amplitudes $(\Psi(Z),W(Z),\Theta(Z))$ are arbitrary, however, and in this sense these solutions are considered `strongly nonlinear' in the terminology of \cite{aB94} since order one distortions of the mean temperature are obtainable with increasing $\Rat$.  The resulting problem consists of a system of ordinary differential equations for the amplitudes $(\Psi(Z),W(Z),\Theta(Z))$, the mean temperature $\mth$ and the nonlinear oscillation frequency $\omega$.  

\begin{figure}
  \begin{center}
   \subfloat[]{
      \includegraphics[height=4.5cm]{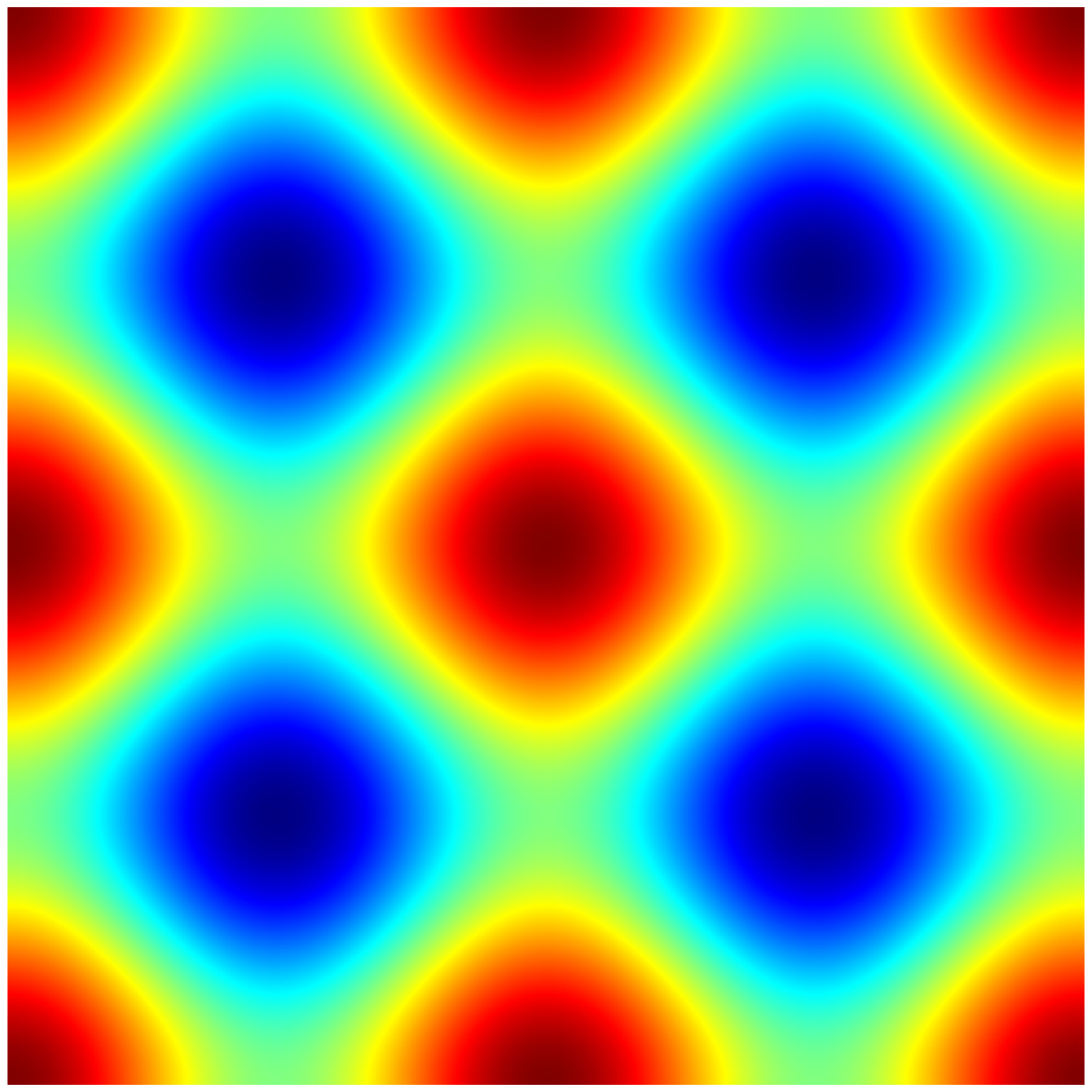}}
      \quad
    \subfloat[]{
      \includegraphics[height=5cm]{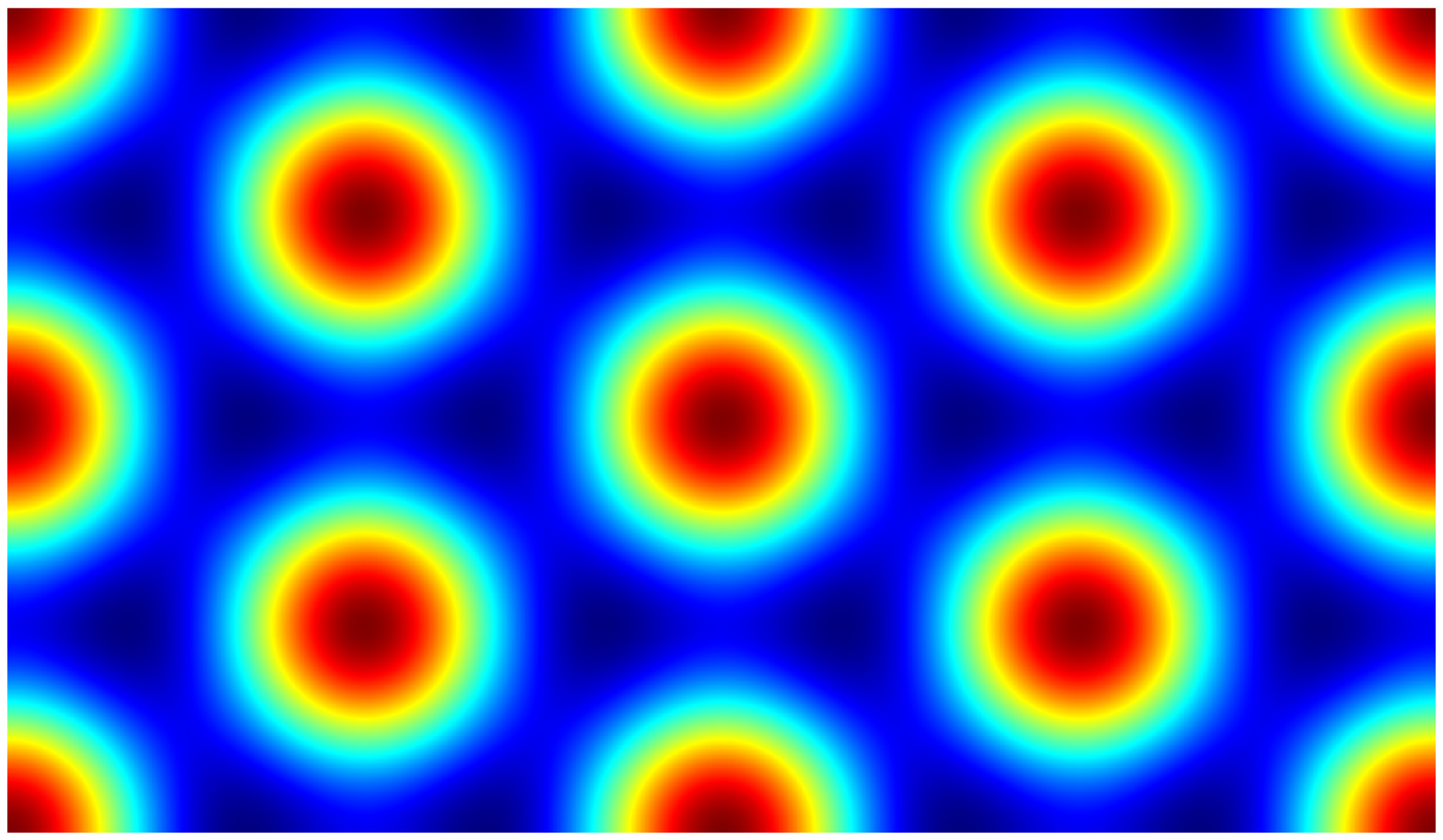}} \\
        \subfloat[]{
      \includegraphics[height=4.5cm]{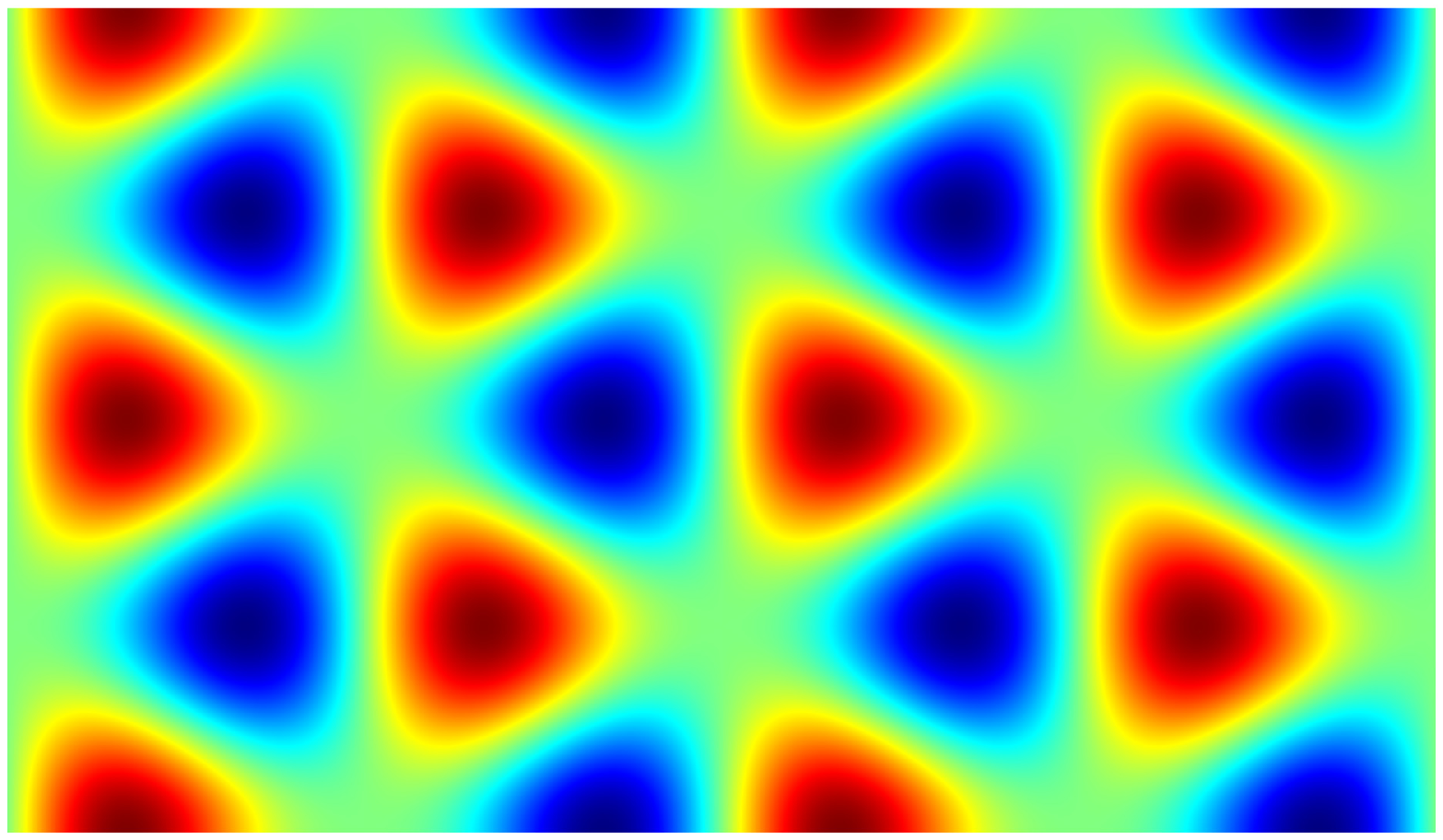}}
      \quad
    \subfloat[]{
      \includegraphics[height=4.5cm]{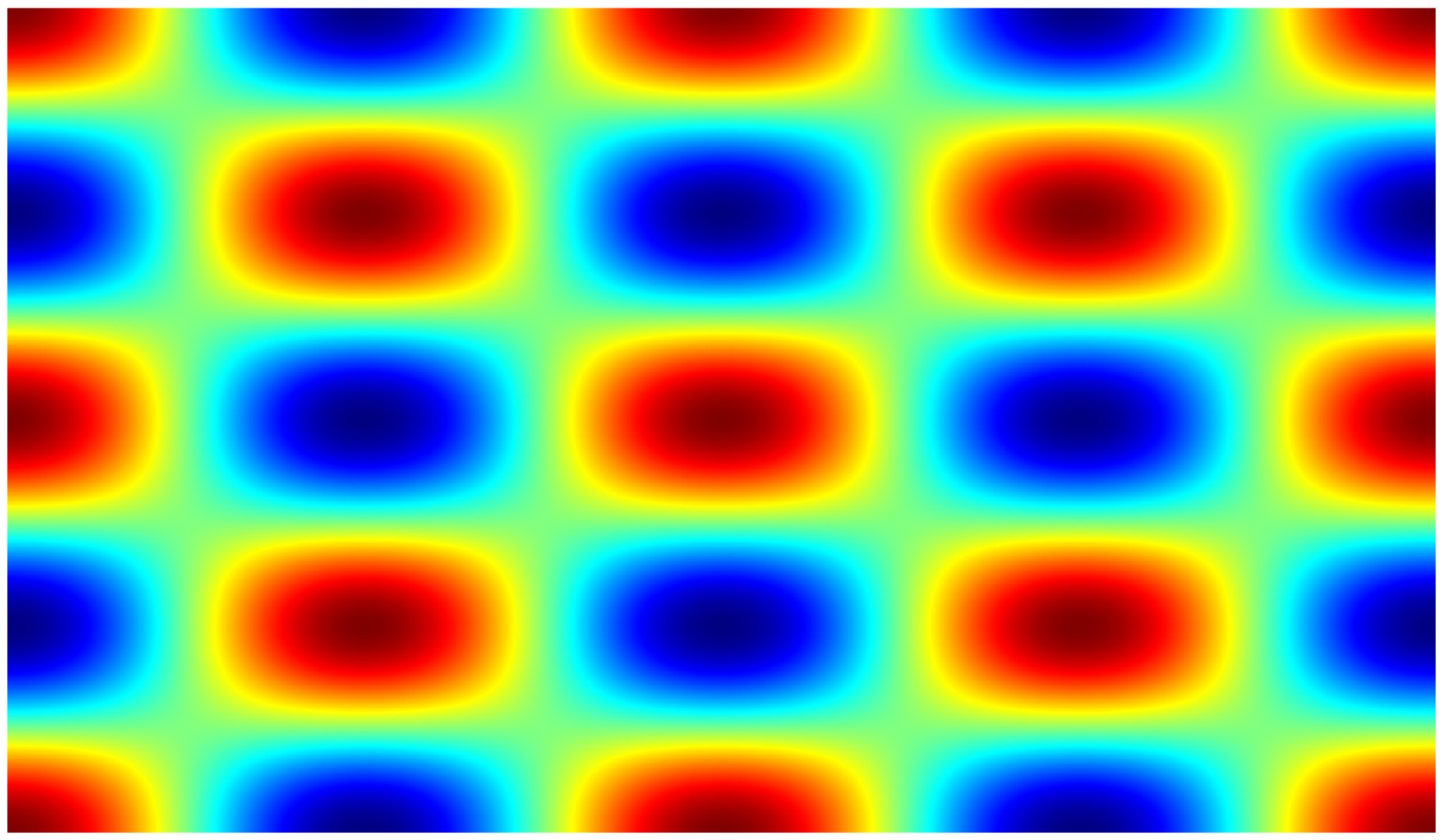}}
  \end{center}
\caption{Visualization of different horizontal $(x,y)$ planforms: (a) squares; (b) hexagons; (c) triangles; and (d) patchwork quilt.  In all cases two wavelengths are shown in each dimension.}
\label{F:plan}
\end{figure}

Results are presented for horizontal planforms consisting of squares, hexagons, triangles, and the patchwork quilt defined by, respectively,
\begin{gather}
h = \cos k x + \cos k y , \label{E:plan1} \\
h = \cos k x + \cos \lb \frac{1}{2} k x + \frac{\sqrt{3}}{2} k y \rb + \cos \lb \frac{1}{2} k x - \frac{\sqrt{3}}{2} k y \rb , \\
h = \sin k x + \sin \lb \frac{1}{2} k x + \frac{\sqrt{3}}{2} k y \rb + \sin \lb \frac{1}{2} k x - \frac{\sqrt{3}}{2} k y \rb , \\
h = \cos \lb \frac{1}{2} k x + \frac{\sqrt{3}}{2} k y \rb + \cos \lb \frac{1}{2} k x - \frac{\sqrt{3}}{2} k y \rb . \label{E:plan4}
\end{gather}
Figure \ref{F:plan} shows the structure of each planform in the horizontal $(x,y)$ plane \citep[see also][]{kJ99b}.  All of the above planforms are characterized by identical linear stability behavior for the onset of convection; the critical Rayleigh number and critical wavenumber for steady convection (independent of $Pr$) , for instance, are $\Rat_c\approx8.6956$ and $k_c\approx1.3048$, respectively.  For $Pr \lesssim 0.68$, convection is oscillatory and the critical parameters depend on $Pr$ \citep[e.g.][]{sC61}.


We note that the horizontal wavenumber $k$ is an undetermined parameter in the single mode theory.  In the following sections we present results that rely on two different approaches for determining the value of $k$.  In the first approach, we simply set $k=k_c$ and calculate the single mode solutions accordingly.  In the second approach, we find the value of $k$ that yields the maximum Nusselt number; a result of this approach is that the wavenumber increases modestly with the Rayleigh number.  For instance, as the solutions are computed the wavenumber changes from $k=1.3048$ at $\Rat=\Rat_c$ up to $k \approx 3$ at $\Rat \approx 500$; hereafter we find it useful to use the notation $Nu_{max}$ when referring to such solutions.

\subsection{Numerical methods}

Numerical solutions to two different mathematical problems were needed in the present work.  The single mode solutions were computed as a nonlinear eigenvalue problem via a Newton-Raphson-Kantorovich (NRK) two-point boundary value solver \citep{pH62,cS82}.  Upon solving for the single mode solutions utilizing the NRK scheme, the solutions $\Psi$ and $W$ were provided as input to the mean induction equations for a given value of $\Rat$.  The resulting generalized eigenvalue problem for the complex eigenvalue $\sigma$ and the mean magnetic field $\mBb$ was solved utilizing MATLAB's `sptarn' function.  Chebyshev polynomials were used to discretize the vertical derivatives appearing in the governing equations.  Up to 800 Chebyshev polynomials were employed to resolve numerically the most extreme cases (i.e., large values of $\Rat$).  To generate a numerically sparse system, we use the Chebyshev three-term recurrence relation and solve directly for the spectral coefficients with the boundary conditions enforced via `tau'-lines \citep{dG93}.  The non-constant coefficient terms appearing in the mean induction equations are treated efficiently by employing standard convolution operations for the Chebyshev polynomials \citep{gB97,sO13}.  An identical approach was used recently by \citep{mC13,mC15,mC15d}.


\section{Results}
\label{S:results}

\subsection{Preliminaries}

In the present section we discuss the salient features of both the QGDM and the single mode solutions.  In particular, we show explicitly the form of the mean induction equation for steady single mode solutions since this is revealing in its mathematical simplicity.  The two components of the mean induction equation \eqref{E:minduc1} are given by
\be
\partial_\tau \mBx = - \dz \emfy  + \frac{1}{\Pmt} \dzt \mBx , \label{E:minduc1a}
\ee
\be
\partial_\tau \mBy = \dz \emfx + \frac{1}{\Pmt} \dzt \mBy , \label{E:minduc1b}
\ee
where 
\be
\emfx = \overline{\lb v' \pbz - w' \pby \rb} , \quad \emfy = \overline{\lb w' \pbx - u' \pbz \rb} .
\ee
With the use of \eqref{E:finduc1} we can eliminate $\bbp$ and the two components of the emf become
\begin{gather}
\emfx = \Pmt \lsq \mBx \lb \overline{w' \lpi \dsxx \psi' - \dsx \psi' \lpi \dsx w'}\rb  +  \mBy \lb \overline{w' \lpi \dsxy \psi' - \dsx \psi' \lpi \dsy w'} \rb \rsq , \label{E:emfx} \\
 \emfy = \Pmt \lsq \mBx \lb \overline{w' \lpi \dsxy \psi' - \dsy \psi' \lpi \dsx w'}\rb  +  \mBy \lb \overline{w' \lpi \dsyy \psi' - \dsy \psi' \lpi \dsy w'} \rb \rsq , \label{E:emfy}
\end{gather}
where $\lpi$ is the inverse horizontal Laplacian operator and is simply $\lpi = -k^{-2}$ for single mode solutions.  We note that in the present kinematic investigation $w'$ and $\psi'$ are computed numerically via the NRK algorithm, and are therefore known a priori.  With the above formulation we can write $\emf_i = \alpha_{ij} \overline{B}_j$, where the components of the `alpha' pseudo-tensor are given by
\be
\alpha_{ij} = \Pmt \lb 
\begin{array}{cc} 
\overline{w' \lpi \dsxx \psi' - \dsx \psi' \lpi \dsx w'}  \, & \,  \overline{w' \lpi \dsxy \psi' - \dsx \psi' \lpi \dsy w'} \\
\overline{w' \lpi \dsxy \psi' - \dsy \psi' \lpi \dsx w'}  &  \overline{w' \lpi \dsyy \psi' - \dsy \psi' \lpi \dsy w'}
\end{array}
\rb . \label{E:alpha}
\ee
The above representation shows that the QGDM allows for an asymptotically exact form of the $\alpha$-effect \citep[e.g.~see][]{hM78b}.  For the particular case of single mode solutions, equations \eqref{E:emfx} and \eqref{E:emfy} simplify to
\be
\emfx = \frac{2 \Pmt}{k^2} \, \overline{(\dsx h)^2} \, \Psi W \mBx , \quad
\emfy = \frac{2 \Pmt}{k^2} \, \overline{(\dsy h)^2} \, \Psi W \mBy .  \label{E:smemf}
\ee
The above result follows from the fact that $\overline{\dsx h \dsy h} \equiv 0$ for the single mode solutions; the consequence is that the alpha tensor becomes diagonal and symmetric,
\be
\alpha_{ij} = \frac{2 \Pmt}{k^2}  \lb 
\begin{array}{cc} 
\overline{(\dsx h)^2} \, \Psi W   \, & \,  0 \\
0  &  \overline{(\dsy h)^2} \, \Psi W
\end{array}
\rb . \label{E:alphasm}
\ee
One might also presume that $\alpha_{ij}$ is symmetric for the more general multi-mode case too since the resulting flows are likely isotropic in the horizontal dimensions, though symmetry breaking effects such as a tilt of the rotation axis may change this property \citep[e.g.][]{kJ98,fP02}.  

The form of the emf given in equations \eqref{E:smemf} illustrates one of the primary differences with the weakly nonlinear investigations of \citep{aS74,kM13}, and the linear  investigation of \citep{bF13}.  Denoting the critical Rayleigh number for convection as $\Rat_c$, these previous works considered either asymptotically small deviations (i.e.~$\Rat \approx \Rat_c$) from linear convection \citep{aS74,kM13}, or no deviation ($\Rat \equiv \Rat_c$) in the case of \citep{bF13}.  In the present work the convection is strongly forced and thus $W$ and $\Psi$ exhibit strong departures from their corresponding linear functional forms $W\sim \sin(\pi Z)$ and $\Psi \sim \cos(\pi Z)$.  Moreover, here we consider the case of asymptotically small magnetic Prandtl number, rather than the $Pm=O(1)$ case investigated in prior work.

Equations \eqref{E:smemf} show that for the case of rolls oriented in either the $x$ or $y$ direction dynamo action is not possible since coupling between the two components of the mean induction equation is lost.  For instance, rolls oriented in the $y$-direction are given by $h = \cos k x$ and thus $\emfy \equiv 0$.  The invariance of equation \eqref{E:alpha} under arbitrary rotations in the horizontal plane implies that this result holds for all roll orientations.  In contrast, dynamos exist for the four planforms defined by equations \eqref{E:plan1}-\eqref{E:plan4}.

The averages appearing in equations \eqref{E:smemf} take the form $\overline{(\dsx h)^2} = c_1 k^2$  and $\overline{(\dsy h)^2} = c_2 k^2$, where $c_1$ and $c_2$ are numerical prefactors that depend upon the planform employed.  For squares we have $c_1 = c_2 = \frac{1}{2}$, hexagons and triangles yield $c_1 = c_2 = \frac{3}{4}$, whereas the patchwork quilt gives $c_1 = \frac{3}{4}$ and $c_2 = \frac{1}{4}$.  The onset of dynamo action is therefore solely dependent upon these constants, the magnetic Prandtl number and the Rayleigh number.   We note that $\alpha_{11}=\alpha_{22}$ when $c_1=c_2$, as it is for all planforms considered here except the patchwork quilt.

The mean kinetic helicity, defined by $\He = \overline{\ubp \cdot \boldsymbol{\zeta}'}$, plays an important role in the dynamo mechanism \citep[e.g.][]{hM78b}.  Utilizing the reduced vorticity vector 
\be
\boldsymbol{\zeta} = \dsy w' \, \hx - \dsx w' \, \hy + \lb \dsx v' - \dsy u' \rb \hz,
\ee
the mean helicity becomes
\be
\He = 2 \, \overline{w' \lp \psi'} .
\ee
For single mode solutions this becomes $\He = - 2 k^2 \Psi W,$ showing a direct correspondence between the components of the single mode alpha tensor \eqref{E:alphasm} and helicity.  

We also find it useful to examine the relative helicity, defined by
\be
\He_r = \frac{\overline{\ubp \cdot \boldsymbol{\zeta}'}}{ \overline{\ub^{\prime2}}^{1/2} \, \overline{\boldsymbol{\zeta}^{\prime2}}^{1/2} } = \frac{- 2 k \Psi W}{ k^2 \Psi^2 + W^2}  ,
\ee
and we note that $\He_r$ is independent of the particular planform employed.

\begin{figure}
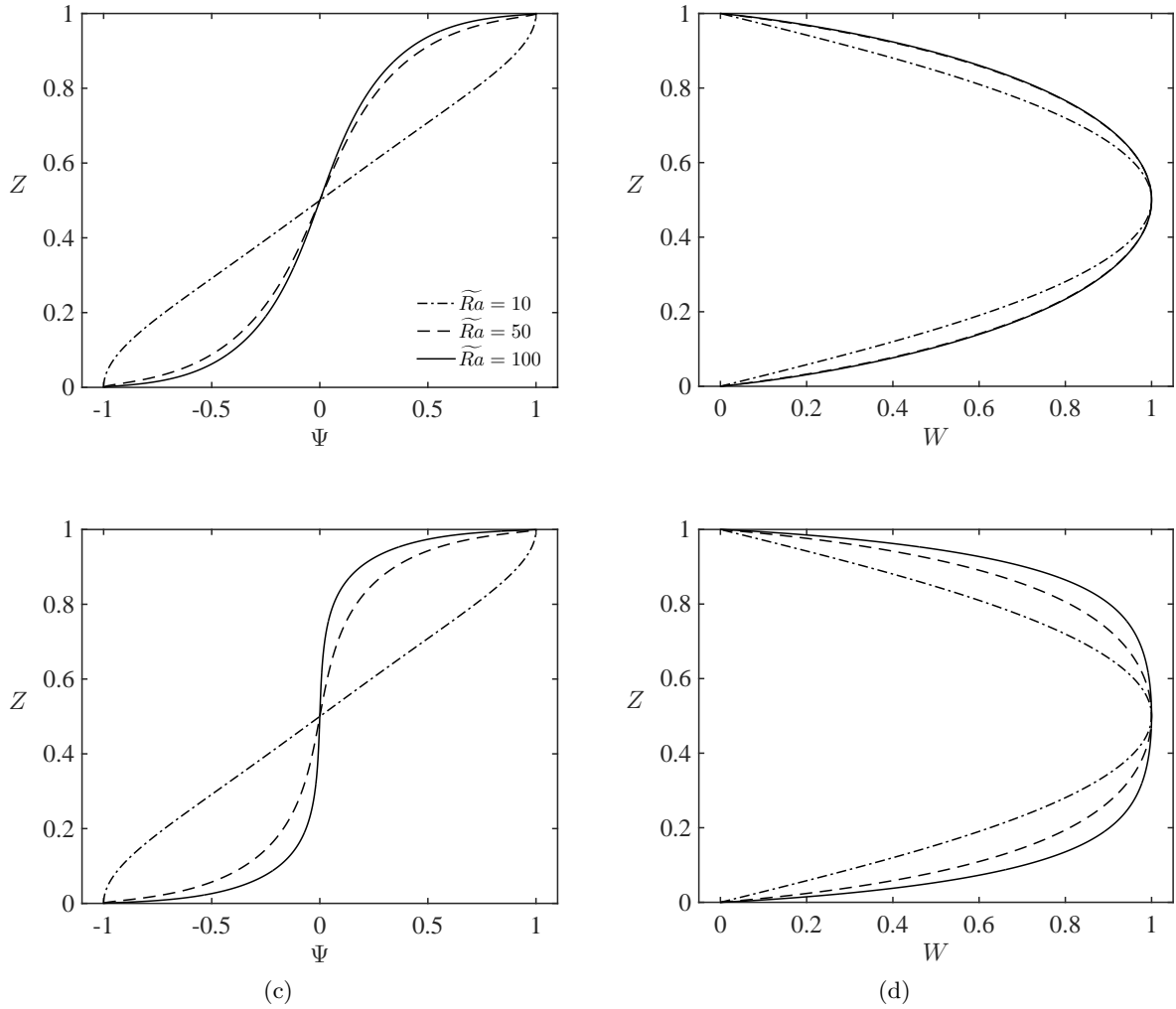

  \begin{center}
      \subfloat[]{
      \includegraphics[height=6cm]{Fig2a}}
      \qquad
    \subfloat[]{
      \includegraphics[height=6cm]{Fig2b}} \\      
   \subfloat[]{
      \includegraphics[height=6cm]{Fig2c}}
      \qquad
   \subfloat[]{
      \includegraphics[height=6cm]{Fig2d}}     
  \end{center}
\caption{Normalized vertical profiles for steady convection and both classes of single mode solutions:  (a),(c) stream function ($\Psi$); and (b),(d) vertical velocity ($W$).  Plots (a) and (b) correspond to the fixed wavenumber case with $k=1.3048$ and plots (c) and (d) correspond to the maximum Nusselt number case. }
\label{F:smodes1}
\end{figure}

\begin{figure}
  \begin{center}
    \subfloat[]{
      \includegraphics[height=6cm]{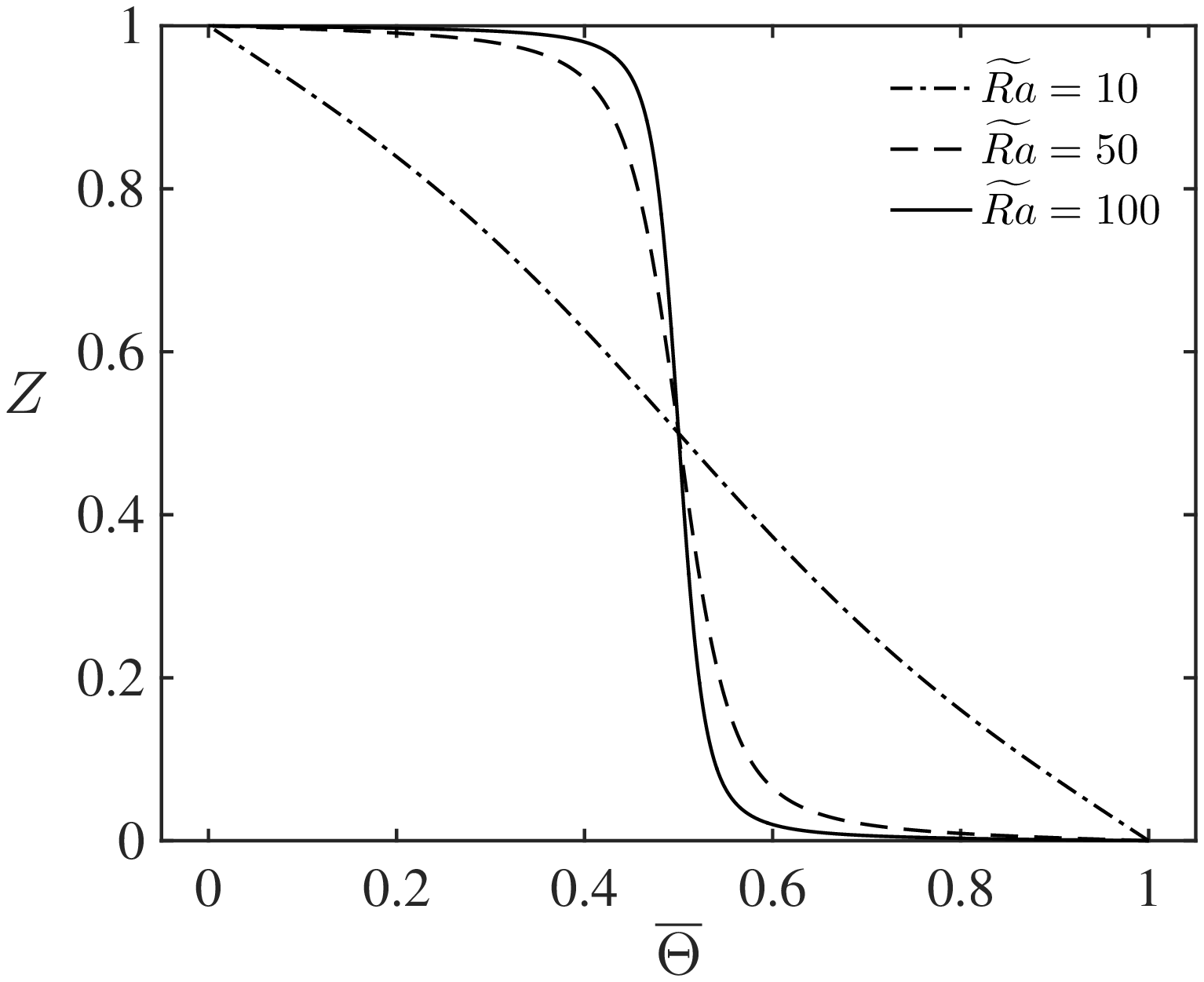}}      
      \qquad
   \subfloat[]{
      \includegraphics[height=6cm]{Fig3b}}     
  \end{center}
\caption{Mean temperature ($\mth$) profiles for steady convection with (a) fixed wavenumber ($k=1.3048$) and (b) maximum Nusselt number single mode solutions for $\Rat = 10, 50, 100$.}
\label{F:smodes2}
\end{figure}

Single mode solutions are shown in Figures \ref{F:smodes1} and \ref{F:smodes2} for $\Rat=10,50$ and $100$.  We limit the present discussion to steady convection since the solutions are qualitatively similar for oscillatory convection.  Steady single mode solutions are Prandtl independent since $Pr$ can be scaled out of the equations.  The $\Rat=10$ case is very similar in structure to the linear eigenfunctions for $\Rat_c=8.6956$.  The structure of the corresponding solutions for the fixed wavenumber case with $k=1.3048$ have been given many times previously, but are included here for the sake of comparison with the $Nu_{max}$ case \cite[e.g.][]{aB94,kJ99b}.  Recent work employed the $Nu_{max}$ case for examining the role of no-slip boundaries on single mode convection, but details on the solution structure were omitted for brevity \citep{kJ15} (see also \citep{iG15}).  Significant differences in the spatial structure for the two cases are apparent and play a role in determining the corresponding dynamo behavior discussed in the next section. For the $k=1.3048$ case, we observe that the shape of the $(\Psi,W)$ fields asymptote and $\dz \mth \propto \Rat^{-1} \rightarrow 0$ in the interior as $\Rat$ is increased \citep[see][]{kJ99b}.  In contrast, for the $Nu_{max}$ case the normalized $(\Psi,W)$ profiles do not possess a recognizable asymptotic structure, yet we find that $
\dz \mth$ does saturate as $\Rat \rightarrow \infty$.  Further, we observe that both $\dz \Psi \rightarrow 0$ and $\dz W \rightarrow 0$ in the fluid interior with increasing $\Rat$.  These observations are in good agreement with nonlinear simulations of equations \eqref{E:fvort2}-\eqref{E:mheat} for $Pr > 1$ \citep[e.g.][]{kJ12}; this agreement is likely due to the decreased role of advective nonlinearities in large $Pr$ convection.

\begin{figure}
  \begin{center}
      \subfloat[]{
      \includegraphics[height=6cm]{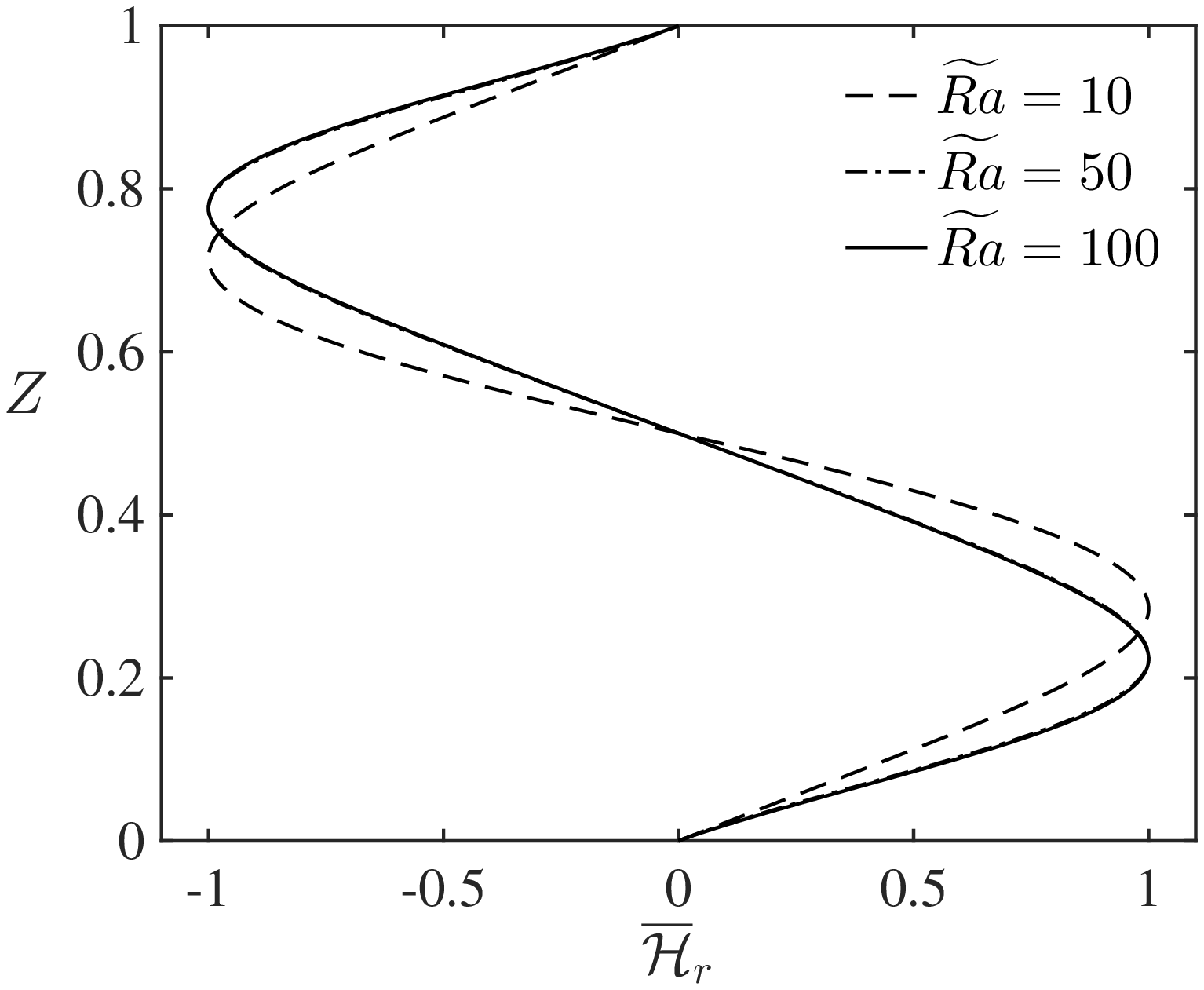}}
      \qquad
    \subfloat[]{
      \includegraphics[height=6cm]{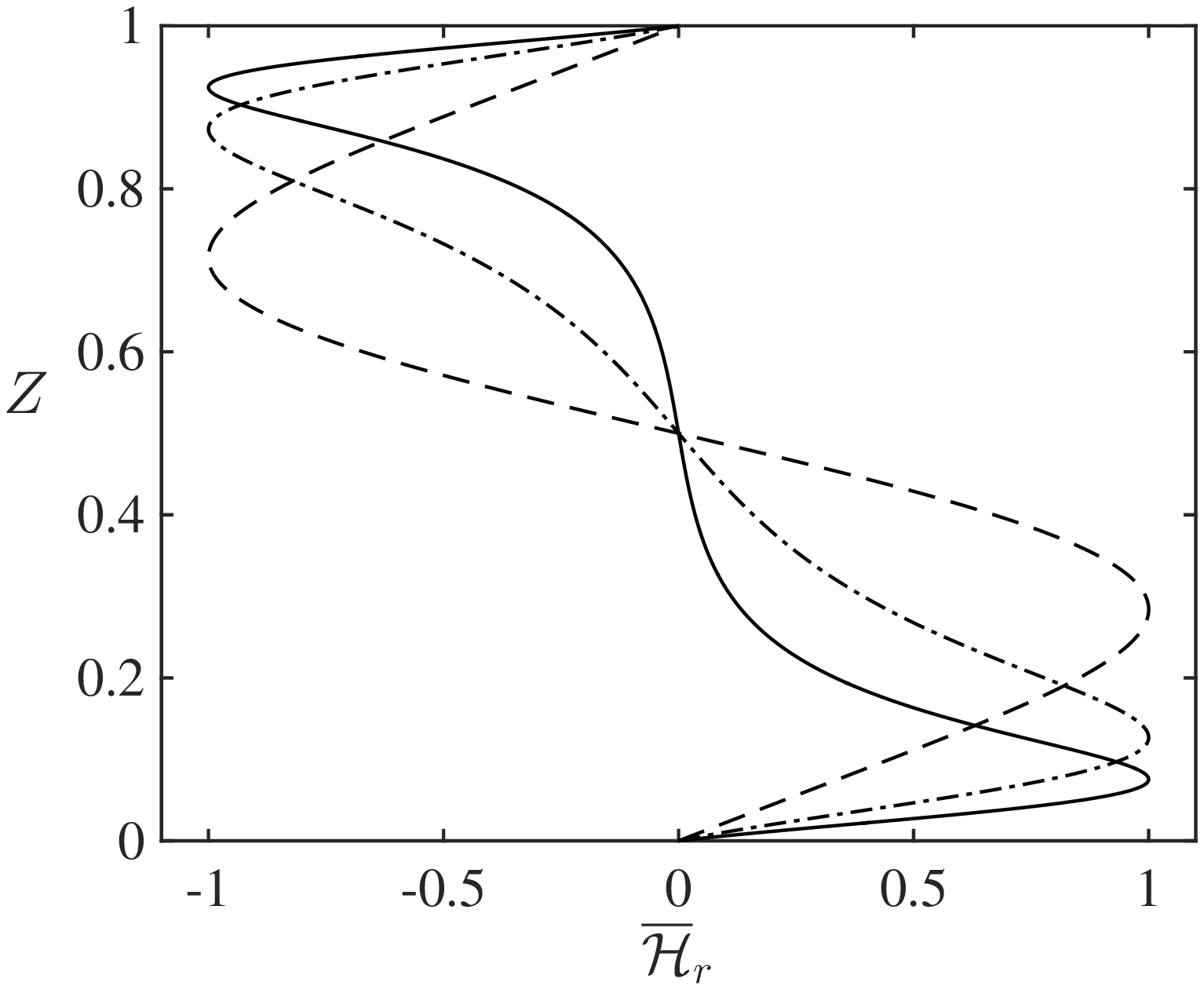}} \\      
  \end{center}
\caption{Relative helicity profiles for steady convection with $\Rat=10, 50$ and $100$ and both (a) fixed wavenumber single mode solutions and (b) maximum Nusselt number single mode solutions. Note that the normalized profiles asymptote as $\Rat \rightarrow \infty$ when $k$ is fixed (the profiles for $\Rat=50$ and $100$ are nearly indistinguishable), whereas the profiles in (b) do not asymptote and exhibit boundary layer behavior.}
\label{F:hel}
\end{figure}

\begin{figure}
  \begin{center}
      \includegraphics[height=7cm]{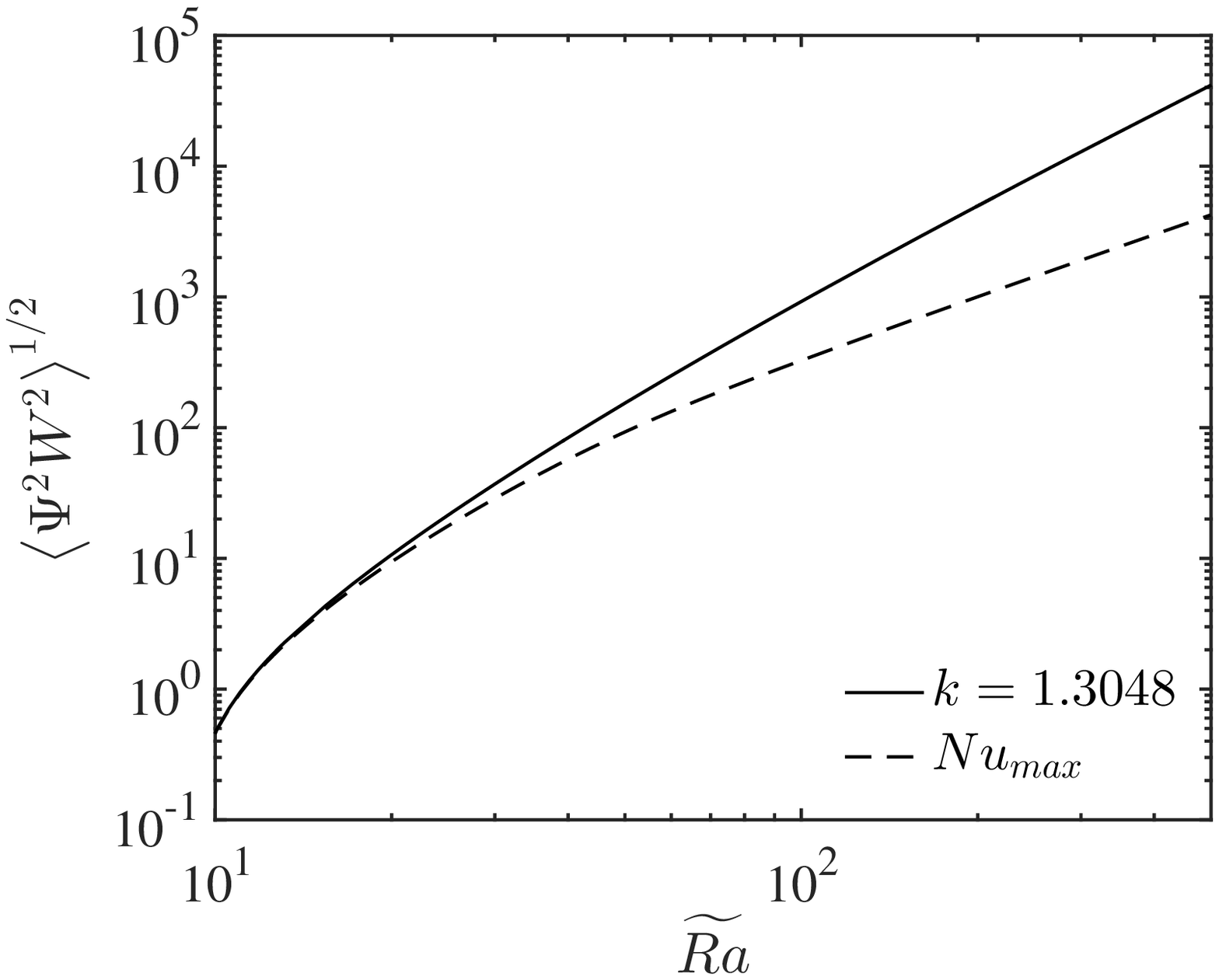}  
  \end{center}
\caption{The root-mean-square of the product $\Psi W$ (proportional to the kinetic helicity) is shown as a function of $\Rat$ for both classes of single mode solutions. }
\label{F:hel2}
\end{figure}

Figure \ref{F:hel} illustrates the $\Rat$ dependence of the relative helicity $\He_r$ for both the fixed-$k$ and $Nu_{max}$ single mode solutions.  For the $k=1.3048$ case (Figure \ref{F:hel}a) the profiles asymptote as $\Rat \rightarrow \infty$.  In contrast for the $Nu_{max}$ case (Figure \ref{F:hel}b) we do not observe asymptotic behavior of $\He_r$ as $\Rat$ becomes large.  Moreover, the $\He_r$ profiles for the $Nu_{max}$ case show boundary layer behavior near the top and bottom boundaries.  Figure \ref{F:hel2} shows the root-mean-square (rms) helicity, $\langle \Psi^2 W^2 \rangle^{1/2}$, as a function of $\Rat$ for both classes of solutions; for both cases the rms of the product $\Psi W$ increases monotonically as a function of $\Rat$ with the increase observed to be more rapid for the fixed-$k$ case.


\subsection{Kinematic dynamos}

The single mode solutions discussed in the preceding section are now employed for determining the onset of dynamo action.  For a given value of the Rayleigh number and associated single mode solutions, we compute the value of the magnetic Prandtl number, denoted $\Pmt_d$, that yields a magnetic field with zero growth rate (i.e.~marginal stability) by solving the generalized eigenvalue problem.  As the Reynolds number is fixed by $\Rat$, this is equivalent to determining the critical magnetic Reynolds number.  The oscillation frequency of the dynamo, referred to as the dynamo frequency, is denoted by $\sigma_d$; we stress that this oscillation occurs on the intermediate mean magnetic field timescale $\tau$.

\begin{figure}
  \begin{center}
   \subfloat[]{
      \includegraphics[height=6.2cm]{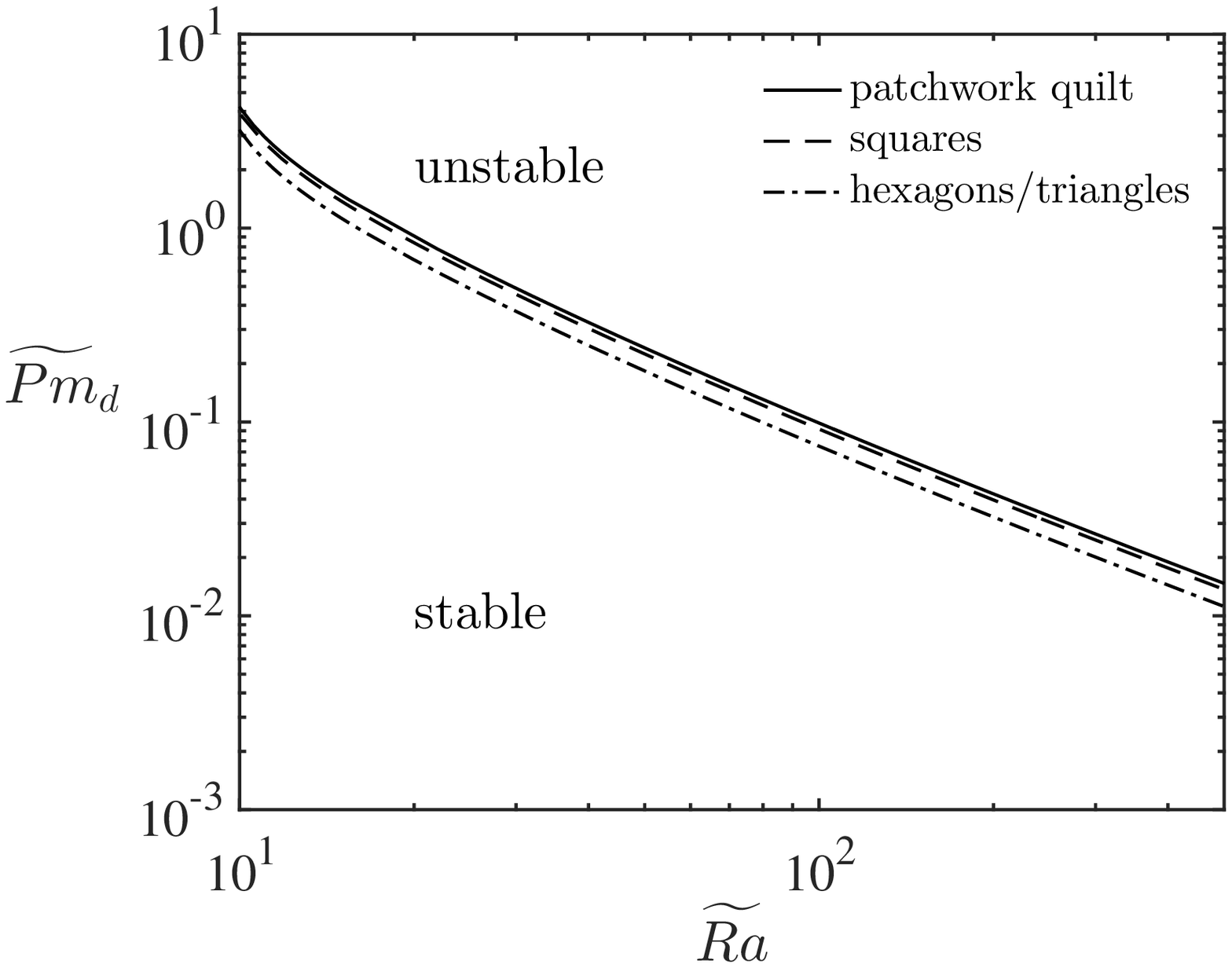}}
      \qquad
    \subfloat[]{
      \includegraphics[height=6.2cm]{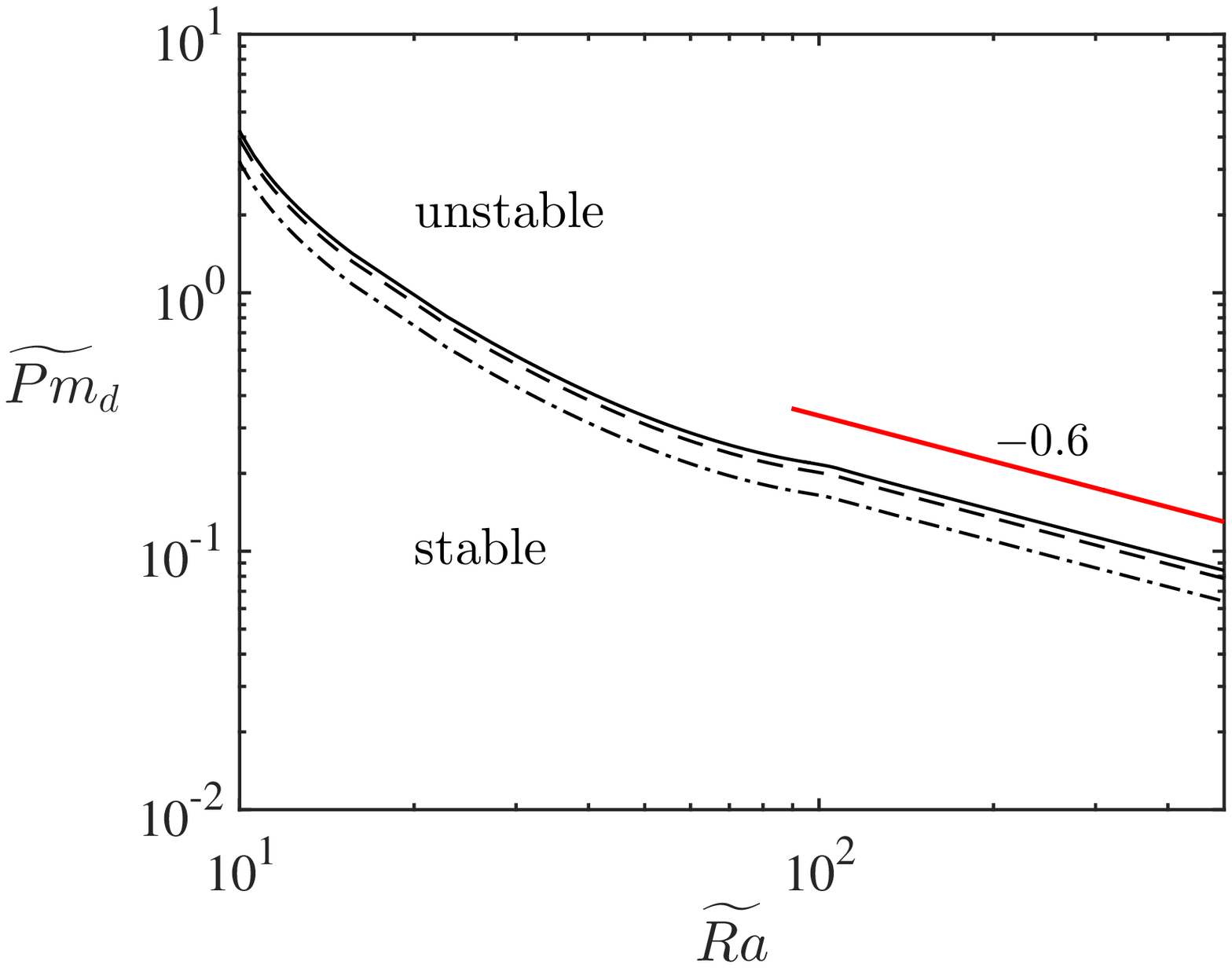}} \\
  \end{center}
\caption{Kinematic dynamo results for steady convection with both classes of single mode solutions and the various horizontal planforms defined by equations \eqref{E:plan1}-\eqref{E:plan4}.  Results are shown for (a) fixed wavenumber solutions and (b) maximum Nusselt number solutions. The red solid line in (b) show the logarithmic slope of the numerically-fitted power law scaling behavior $\Pmt_d \sim \Rat^{-0.6}$.}
\label{F:kin}
\end{figure}

\begin{figure}
  \begin{center}
   \subfloat[]{
      \includegraphics[height=6.2cm]{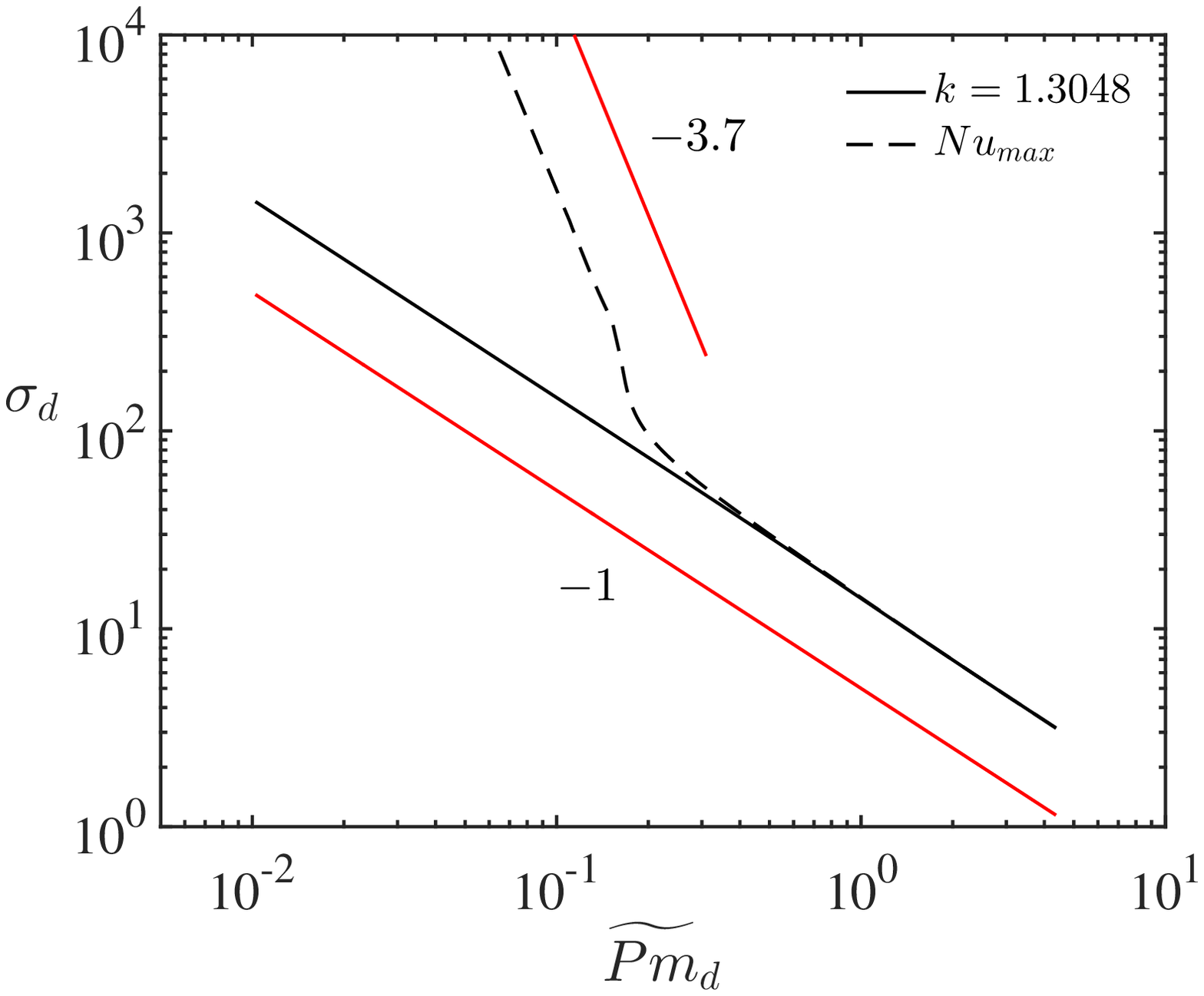}}
      \qquad
    \subfloat[]{
      \includegraphics[height=6.2cm]{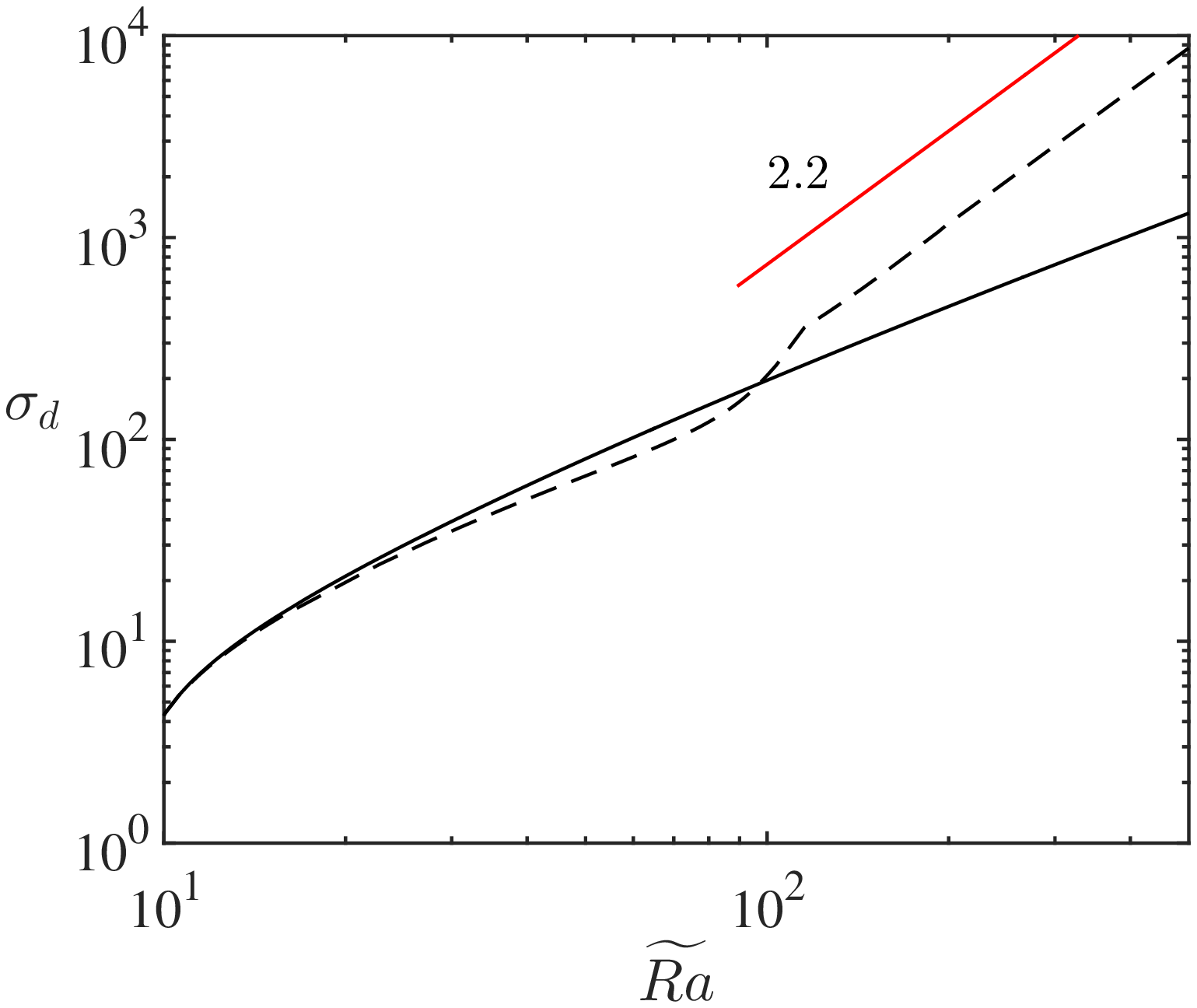}}                
  \end{center}
\caption{Critical dynamo frequency $\sigma_d$ versus (a) $\Pmt_d$ and (b) $\Rat$ for both classes of single mode solutions. Only results for steady convection with hexagonal and triangular planform are shown.  The red solid lines show the logarithmic slopes of the numerically-fitted power law scaling behavior, i.e.~$\sigma_d \sim \Pmt_d^{-1}$ and $\sigma_d \sim \Pmt_d^{-3.7}$ in (a) and $\sigma_d \sim \Rat^{2.2}$ in (b).}
\label{F:sig}
\end{figure}

Figure \ref{F:kin} shows results from the computations both for (a) fixed wavenumber and (b) $Nu_{max}$ cases for steady convection ($Pr \gtrsim 0.68$).  All of the computed magnetic field solutions are oscillatory, with the $\Pmt_d$ and $\Rat$ dependence of $\sigma_d$ shown in Figure \ref{F:sig}.  Hexagons and triangles have identical dynamo behavior given that they are characterized by the same averaging constants $c_1$ and $c_2$.  Moreover, for a given value of $\Pmt_d$, these two planforms are always characterized by the lowest value of $\Rat$ for dynamo action since the associated averages, as quantified by $c_1$ and $c_2$, are largest for these planforms.  In contrast, the patchwork quilt requires the largest values of $\Pmt$ for a given $\Rat$ and suggests that anisotropy in $\alpha$ reduces the efficiency of dynamo action.  Both classes of single mode solutions show a monotonic decrease in $\Pmt_d$ as $\Rat$ is increased.  However, the $Nu_{max}$ case shows a noticeable change in dynamo behavior near $\Pmt_d \approx 0.2$, corresponding to $\Rat \approx 100$.  As shown in Figure \ref{F:hel}(b), boundary layer behavior occurs in the single mode solutions and this behavior becomes most pronounced for $\Rat \gtrsim 100$. In Figure \ref{F:kin}(b) we show the logarithmic slope of the numerically-fitted power law scaling $\Pmt_d \approx 2.5 \Rat^{-0.6}$.


\begin{figure}
  \begin{center}
      \includegraphics[height=7cm]{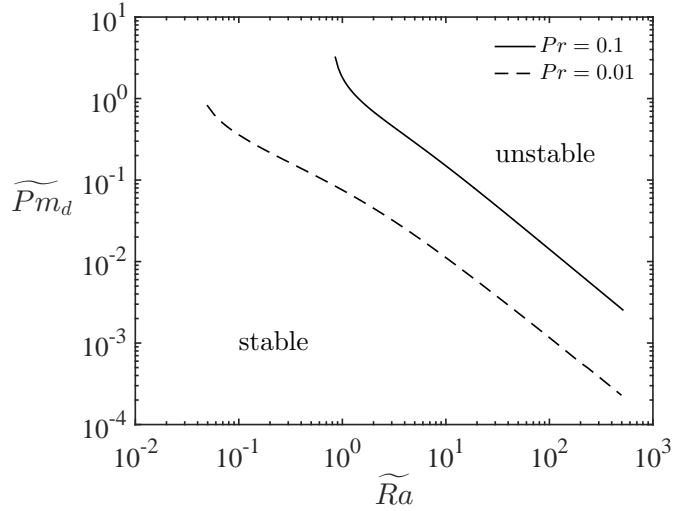}        
  \end{center}
\caption{Kinematic dynamo results for oscillatory convection with $Pr=0.1$ (solid curve) and $Pr=0.01$ (dashed curve).  For simplicity only the results for hexagonal and triangular planforms are shown; other planforms have qualitatively similar results.  The wavenumber was fixed at the critical wavenumber for each respective case.}
\label{F:kinLowPr}
\end{figure}

Figure \ref{F:kinLowPr} displays results for oscillatory convection at $Pr=0.1$ and $Pr=0.01$; the critical parameters for these respective values are ($\Rat_c$, $k_c$, $\omega_c$) = ($0.7820$, $0.5867$, $ 4.8313$) and ($0.03734$, $0.2802$, $11.1009$).  For these calculations the wavenumber was fixed at the critical wavenumber for each Prandtl number.  An interesting consequence of the reduced critical Rayleigh number for low Prandtl numbers is that the convective amplitudes of the oscillatory single mode solutions reach amplitudes comparable with the steady cases at significantly lower values of $\Rat$.  As a result, the low $Pr$ solutions are capable of generating dynamos at much lower critical magnetic Prandtl numbers than the steady solutions ($Pr \gtrsim 1$) for a given value of the Rayleigh number.  For instance, the critical magnetic Prandtl number for $\Rat \approx 10$ is $\Pmt_d \approx 1$ for steady convection, whereas for $Pr=0.01$ dynamo action occurs at $\Pmt_d \approx 10^{-2}$ for a comparable Rayleigh number.  Aside from the slightly different shape in the $\Pmt_d$ vs $\Rat$ marginal curves, the vertical structure of the eigenfunctions and the $\Rat$-dependence of the critical frequency was observed to be nearly the same as that associated with the steady solutions shown in Figure \ref{F:kin}.  For this reason, the discussion hereafter focuses on the solutions obtained from steady convection.


The dependence of dynamo action on $\Pmt$ is due to imbalances in the relative magnitudes of the induction and diffusion terms present within the mean induction equations \eqref{E:minduc1a}-\eqref{E:minduc1b} when $\Pmt \ne 1$.  With $\Pmt > 1$ the influence of magnetic diffusion is reduced and dynamo action can be sustained for a lower value of $\Rat$; the opposite trend holds when $\Pmt < 1$.  Clearly, increasing $\Pmt$ yields larger alpha coefficients (e.g.~\eqref{E:alphasm}); the converse is true when $\Pmt$ is reduced.  

For a given value of $\Rat$ we can obtain an order of magnitude estimate for the value of $\Pmt_d$ by balancing induction with diffusion in the mean induction equations \eqref{E:minduc1a}-\eqref{E:minduc1b},
\be
\Pmt_d \sim |\Psi W |^{-1/2}, \label{E:pmt}
\ee
where we have assumed $\dz = O(1)$.  Thus, the rms helicity shown in Figure \ref{F:hel2} provides a direct estimate for $\Pmt_d$.  


%
%

The dependence of $\sigma_d$ on $\Pmt_d$ shown in Figure \ref{F:sig}(a) can be understood by considering the linearity of the mean induction equations \eqref{E:minduc1a}-\eqref{E:minduc1b}.  For a marginally stable oscillatory dynamo to exist, we must have 
\be
\dtau \mBb \sim \frac{1}{\Pmt} \partial_Z^2 \mBb , \label{E:bal1}
\ee
since magnetic diffusion must be present to halt the growth of the field.  It then follows that $\dtau = \sigma_d = O(\Pmt^{-1})$.  This scaling is well demonstrated in Figure \ref{F:sig}(a) for the entire range of $\Rat$ for the fixed wavenumber single mode solutions.  An alternative interpretation of this scaling is to employ relation \eqref{E:pmt} such that $\sigma_d \sim |\Psi W |^{1/2}$.  The dynamo frequency therefore increases with the vigor of convection and thus with $\Rat$, as demonstrated in Figure \ref{F:sig}(b).

For the $Nu_{max}$ solutions we observe the same $\sigma_d \sim \Pmt_d^{-1}$ behavior for $\Rat \lesssim 10^2$ ($\Pmt_d \gtrsim 0.2$) in Figure \ref{F:sig}(a), but for larger values of $\Rat$ (smaller values of $\Pmt_d$) a transition region and subsequent change to a significantly steeper scaling for $\sigma_d$ is observed. From the data we find the two power-law fits $\sigma_d  \approx 0.3 \Pmt_d^{-3.7}$ and $\sigma_d \approx 0.01 \Rat^{2.2}$, with the slopes shown by the red solid lines in Figures \ref{F:sig}(a) and \ref{F:sig}(b), respectively. The boundary layer behavior shown in the helicity plots of Figure \ref{F:hel} shows that vertical derivatives now become large near the top and bottom boundaries so that the assumption $\dz = O(1)$ no longer holds.  If we then assume a power-law scaling of the vertical derivative such that $\dz = O \lb \Rat^\beta \rb$ with $\beta > 0$, the balance expressed in \eqref{E:bal1} then becomes
\be
\sigma_d \sim \frac{\Rat^{2 \beta }}{\Pmt_d} . \label{E:sigNu}
\ee
If we use the empirically determined relation $\Pmt_d \approx 2.5 \Rat^{-0.6}$ shown in Figure \ref{F:kin}(b) to eliminate $\Rat$ we have
\be
\sigma_d \sim \Pmt_d^{-3.4 \beta - 1} , \label{E:sigNu}
\ee
showing that the steep scaling of $\sigma_d$ for the $Nu_{max}$ solutions given in Figure \ref{F:sig}(a) likely follows from enhanced vertical derivatives.

%
%
%
%
%
%

\begin{figure}
  \begin{center}
  \subfloat[]{
      \includegraphics[height=6cm]{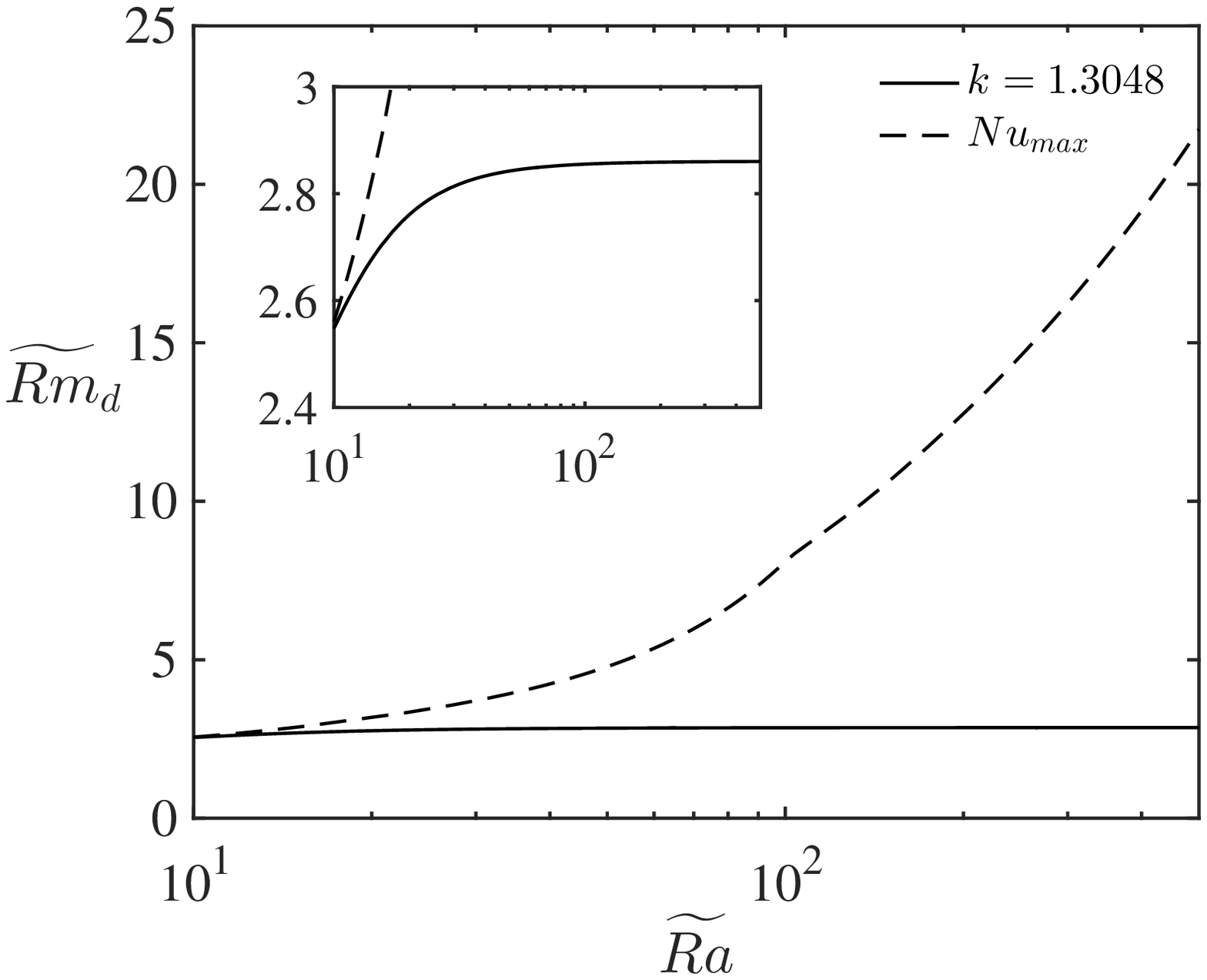}}
      \qquad
   \subfloat[]{        
      \includegraphics[height=6cm]{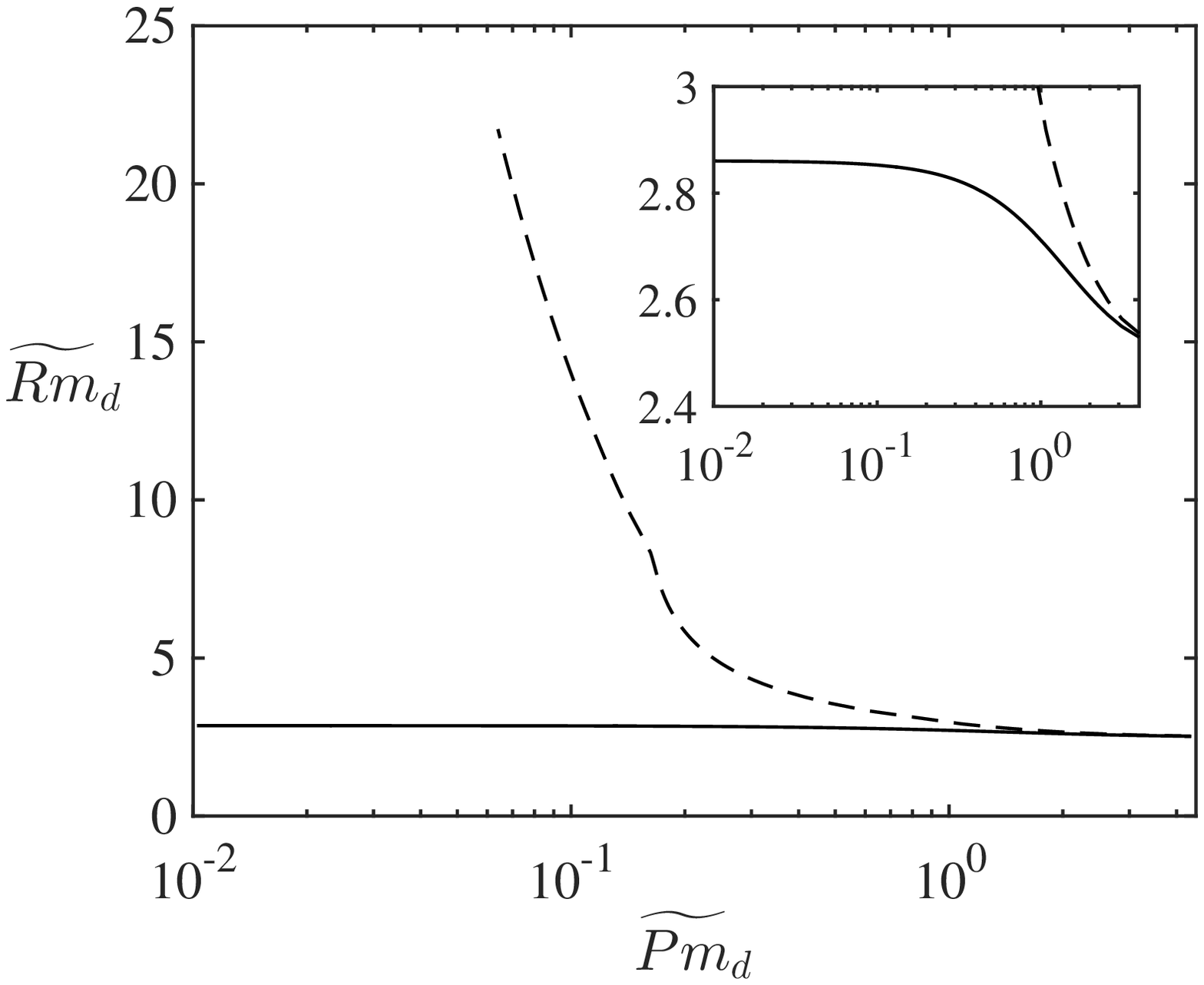}} 
  \end{center}
\caption{Reduced critical magnetic Reynolds number $\Rmt_d = \Pmt_d Re$ versus (a) $\Rat$ and (b) $\Pmt_d$ for steady convection and both classes of single mode solutions.  For the fixed wavenumber case (solid curve) $\Rmt_d \rightarrow \approx 2.9$ as $\Rat \rightarrow \infty$ ($\Pmt_d \rightarrow 0$).  The inset figures show the same data with a reduced ordinate range to highlight the asymptotic behavior of the fixed wavenumber case.}
\label{F:Rm}
\end{figure}

We can also determine the local magnetic Reynolds number required for dynamo action upon noting that $\Rmt = \Pmt Re$.  With the viscous diffusion scaling employed in the present work we take $Re = \langle W^2 \rangle^{1/2}$.  Figure \ref{F:Rm} shows $\Rmt_d$ as a function of (a) $\Rat$ and (b) $\Pmt_d$.  We observe that $\Rmt_d$ asymptotes as the Rayleigh number becomes large (and $\Pmt_d$ becomes small), beginning at a minimum value of $\Rmt_d \approx 2.5$ near $\Rat_c$ and approaching $\Rmt_d \approx 2.9$ as $\Rat$ is increased.  For the $Nu_{max}$ solutions we find that $\Rmt_d$ increases monotonically with $\Rat$; for these solutions $Re \sim \Rat^{1.2}$ for $\Rat \gtrsim 100$ and thus $\Rmt_d \sim \Rat^{0.6}$.  The results show that $Re$ grows at a sufficiently large rate with $\Rat$ to counteract the reduction in $\Pmt_d$ with $\Rat$, demonstrating that boundary layer behavior plays an important role in the growth (or decay) of magnetic field.


\begin{figure}
  \begin{center}
   \subfloat[]{
      \includegraphics[height=8cm]{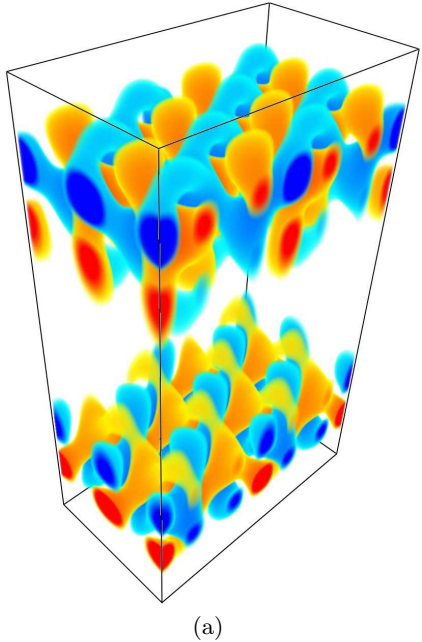}}          
  \end{center}
\caption{Perspective view of the vertical component of the fluctuating magnetic field ($\bz$) for the maximum Nusselt number solution with $\Rat=200$ and triangular planform.  Two convective wavelengths are shown in each horizontal dimension. }
\label{F:bz}
\end{figure}

Figure \ref{F:bz} shows a perspective view of the vertical component of the small-scale magnetic field for the $Nu_{max}$ single mode solution with $\Rat=200$  ($\Pmt = 0.11$) and triangular planform.  The magnetic field is concentrated at vertical depths of $Z \approx 0.2, 0.8$ as a consequence of the depth-dependence of the helicity given in Figure \ref{F:hel}.  The small-scale magnetic field is similar for other values of the Rayleigh number, with the maximum values of the magnetic field moving towards the top and bottom boundaries in accordance with the observed behavior of the helicity and mean magnetic field for these solutions.  Similar behavior of the small-scale magnetic field has been observed in the DNS study of \citep{sS04} (see their Fig. 10).


\begin{figure}
  \begin{center}
      \includegraphics[height=6cm]{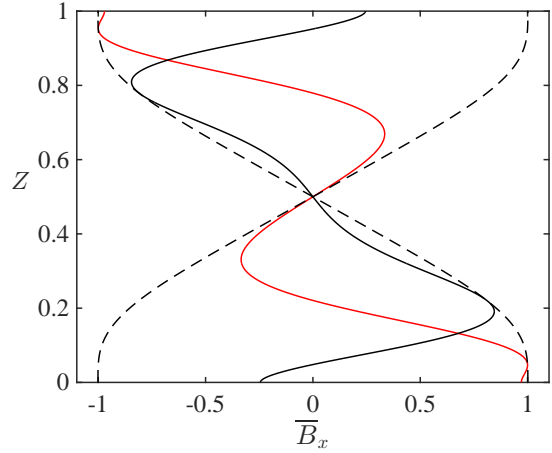}             
  \end{center}
\caption{Mean magnetic field eigenfunctions for the steady fixed wavenumber single mode solutions with $\Rat=50$.  All other Rayleigh numbers show a similar spatial structure for this class of single mode solutions.  The real and imaginary parts of the solutions are shown by the solid black and solid red curves, respectively, and the modulus is given by the dashed black curves.}
\label{F:eig1}
\end{figure}

\begin{figure}
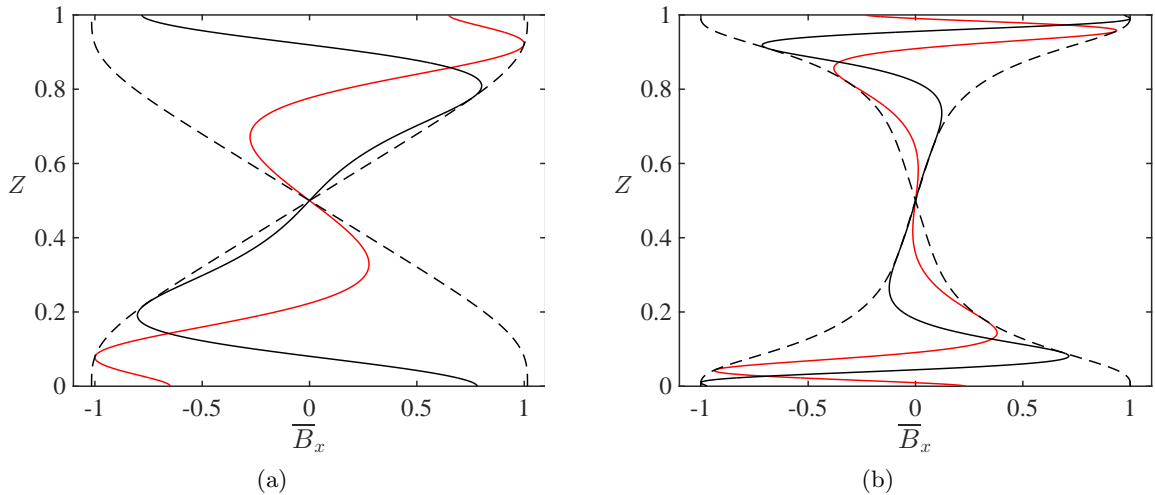

  \begin{center}
   \subfloat[]{
      \includegraphics[height=6cm]{Fig12a}}
      \qquad
    \subfloat[]{
      \includegraphics[height=6cm]{Fig12b}}               
  \end{center}
\caption{Mean magnetic field eigenfunctions for the steady maximum Nusselt number single mode solutions for (a) $\Rat=50$ and (b) $\Rat=200$. All curves are as defined in Figure \ref{F:eig1}.}
\label{F:eig2}
\end{figure}

The $x$-component mean magnetic field eigenfunction is plotted in Figure \ref{F:eig1} for the case of fixed horizontal wavenumber and $\Rat=50$.  The real and imaginary components are shown by the solid black and red curves, respectively, and the eigenfunction envelope $\pm |\mBx|$ is given by the dashed black curves; identical results are found for $\mBy$.   Only a slight $\Rat$-dependence of the spatial structure of $\mBb$ was observed for the fixed wavenumber solutions in the sense that the solution shown in Figure \ref{F:eig1} is representative across the entire range of $\Rat$ investigated.  This result is due to the asymptotic behavior observed in the profiles of helicity plotted in Figure \ref{F:hel}(a).  In contrast, the magnetic field eigenfunctions shown in Figure \ref{F:eig2} for the maximum Nusselt number solutions show a strong $\Rat$-dependence in spatial structure.  For increasing $\Rat$ these magnetic eigenmodes become increasingly localized near the top and bottom boundaries, in agreement with the boundary layer behavior shown in the helicity profiles of Figure \ref{F:hel}b.  We find that the mean magnetic energy then becomes concentrated near the boundaries.


\section{Conclusion}
\label{S:conclude}

The present work investigates the kinematic dynamo problem for finite amplitude, single mode solutions in the context of the low magnetic Prandtl number dynamo model recently developed by\citep{mC15b}, and is the first investigation of dynamo action in the limit of asymptotically small Rossby and magnetic Prandtl numbers.  The single mode solutions allow the kinematic dynamo problem to be solved with relative ease given their simple single-frequency time dependence and vertical structure.  The results presented here provide an initial examination of the temporal and spatial dependence of the magnetic field for future investigations of the self-consistent dynamo in which the Lorentz force is incorporated (as shown in equations \eqref{E:fvort}-\eqref{E:fmom}).  An obvious disadvantage of these solutions is that they are implicitly laminar, with no multi-mode interactions present due to the lack of advective nonlinearities.  Nevertheless, the calculations have shown that small magnetic Prandtl number dynamos are easily attainable within the context of low Rossby number convection.  

Results for both steady and unsteady convection were presented.  We find that, in comparison to steady convection, oscillatory low-$Pr$ convection drives dynamo action for a lower magnetic Prandtl number at a given Rayleigh number owing to the decreased critical Rayleigh number.  In regards to planetary dynamos, these results suggest that the low-$Pm$ dynamos typical of liquid metals may be more easily excited via oscillatory thermal convection than via compositional convection, which tends to be characterized by high Prandtl (or Schmidt) numbers \citep[e.g.][]{mP13}.  Apart from this difference, both steady and unsteady convection yield similar results with respect to magnetic field structure.  

The maximum Nusselt number single mode solutions have shown that dynamo action becomes less efficient when boundary layer behavior is present in the horizontally averaged kinetic helicity, due to the enhancement of magnetic diffusion in these regions.  This decrease in efficiency requires larger Rayleigh numbers to generate a dynamo for a given value of the magnetic Prandtl number, in comparison to the fixed wavenumber single mode solutions in which no boundary layer behavior is observed.  These results may have interesting consequences for fully turbulent convection in which boundary layers are inherent \citep[e.g.][]{kJ12b}.  

We find that the behavior of the critical magnetic Reynolds number depends strongly upon the type of single mode solution.  When the wavenumber of the convection is fixed, the critical magnetic Reynolds number asymptotes to a constant value.  In contrast, for the maximum Nusselt number single mode solutions the critical magnetic Reynolds number increases monotonically, further demonstrating that boundary layer behavior plays a crucial role in the dynamo process.  These results are in contrast to DNS investigations of randomly forced dynamos in triply-periodic domains where an eventual decrease in the critical value of $Rm$ is observed as the flow becomes turbulent \citep{yP05,aI07}.  Of course, significant differences exist between the present model and previous DNS studies, two of the most important of these differences being the presence of strong rotational effects and limitation to flows with a single horizontal length scale in the present work.  Furthermore, the simplicity of the single mode solutions implies that changes in flow regime are impossible in the present work, though the growing importance of boundary layers in one class of single mode solutions does appear to play a similar role.  


Our rapidly rotating dynamos are intrinsically multiscale.  This differs significantly from low $Re_H$ dynamos, in which there is no clear scale separation \citep{uC11,jmA15}.  For the present model, the large-scale magnetic field is generated by the $\alpha^2$ process (e.g.~equation \eqref{E:alpha}) via the averaged effects of the small-scale, fluctuating convective motions.  Even though both $Pm \ll 1$ and $Rm \ll 1$ in the present work, equation \eqref{E:finduc1} shows that the dynamo requires small-scale non-zero magnetic diffusion, and thus horizontal spatial gradients, of the fluctuating magnetic field. In the absence of such gradients there is no coupling between the convection and the mean magnetic field; in this sense the fluctuating magnetic field cannot be ``homogenized" on these scales \citep[cf.][]{uC11}.  


Numerical simulations of the non-magnetic quasi-geostrophic convection equations have identified four primary flow regimes that are present in the geostrophic limit as the Rayleigh number is increased above the critical value \cite{kJ12,dN14}.  These regimes consist of (1) cells, (2) convective Taylor columns, (3) plumes and (4) the geostrophic turbulence regime.  The precise location of these regimes is dependent upon the Prandtl number, with the final geostrophic turbulence regime  characterized by an inverse cascade in which the large-scale flow consists of a depth invariant dipolar vortex \citep{aR14}.  The presence of these regimes has also been confirmed in moderate $Pr$ DNS and laboratory experiments \citep{sS14,jC15,jmA15}.  An interesting avenue of future work is determining the influence that the four primary flow regimes have on the onset of dynamo action and further investigations on the influence of the Prandtl number.  Saturated dynamo states in which the Lorentz force is included are also of obvious interest; this work is currently under investigation and will have further implications for understanding natural dynamos.  

\section*{Acknowledgments}
This work was supported by the National Science Foundation under grants EAR \#1320991 (MAC, KJ and JMA) and EAR CSEDI \#1067944 (KJ,  JMA and PM).  Volumetric rendering was performed with the visualization software VAPOR \citep{jC05,jC07}.

\bibliography{../Dynamobib}

\end{document}